 \documentclass[12pt, preprint]{aastex}
 \usepackage{epsf}

\newcommand{\Msun}{M_{\odot}}
\newcommand{\Lsun}{L_{\odot}}
\newcommand{\pc}{{\rm \, pc}}
\newcommand{\kpc}{{\rm \, kpc}}
\newcommand{\mpc}{{\rm \, Mpc}}
\newcommand{\kmps}{{\rm \, km \, s^{-1}}}
\newcommand{\kmpspkpc}{{\rm \, km \, s^{-1} \, kpc^{-1}}}

\newcommand{\K}{{\rm \, K}}
\newcommand{\kkmps}{{\rm \, K \, km \, s^{-1}}}
\newcommand{\unitcnv}{{\rm cm^{-2}\, [K \, km \, s^{-1}]^{-1}}}
\newcommand{\jykmps}{{\rm \, Jy \, km \, s^{-1}}}

\newcommand{\omgb}{\Omega_{\rm b}}
\newcommand{\vsys}{V_{\rm sys}}
\newcommand{\vlsr}{V_{\rm LSR}}
\newcommand{\Halpha}{{\rm H\alpha}}
\newcommand{\HI}{{\rm HI}}
\newcommand{\NII}{{\rm [N\,II]}}
\newcommand{\jypbm}{{\rm \, Jy \, beam^{-1}}}
\newcommand{\mjypbm}{{\rm \, mJy \, beam^{-1}}}

\slugcomment{ApJ in press}

\shorttitle{GAS DYNAMICS IN NGC 3079}
\shortauthors{Koda et al.}
\received{2002 February 1}
\accepted{2002 March 7}

\begin{document}

\title{
NMA CO ($J=1-0$) OBSERVATIONS OF THE $\Halpha$/RADIO LOBE GALAXY NGC 3079: \\
GAS DYNAMICS IN A WEAK BAR POTENTIAL AND CENTRAL MASSIVE CORE}

\author{Koda, J.\altaffilmark{1,2}, Sofue, Y.\altaffilmark{2},
Kohno, K.\altaffilmark{2}, Nakanishi, H.\altaffilmark{2}, Onodera, S.\altaffilmark{2},
Okumura, S. K.\altaffilmark{3}, and Judith A. Irwin\altaffilmark{4}}
\altaffiltext{1}{JSPS Research Fellow; E-mail: koda@ioa.s.u-tokyo.ac.jp}
\altaffiltext{2}{Institute of Astronomy, University of Tokyo, 
    Mitaka, Tokyo 181-0015, Japan}
\altaffiltext{3}{Nobeyama Radio Observatory, Minamisaku, Nagano,
    384-1305, Japan}
\altaffiltext{4}{Department of Physics, Queen's University, Kingston, Ontario, K7L3N6, Canada}

\begin{abstract}
We present $^{12}$CO (1-0) observations in the central $4.5\kpc$ ($1\arcmin$)
of the $\Halpha$/Radio lobe galaxy NGC 3079 with the Nobeyama Millimeter
Array. The molecular gas shows four components: a main disk,
spiral arms, a nuclear disk, and a nuclear core.
The main disk extends along the galaxy major axis. We detected its central
$2\kpc$ radius, while its full extent is beyond our spatial coverage.
Molecular gas is smoothly distributed in the main disk, having a gas mass
of $5 \times 10^9\Msun$ within the central $\sim 2\kpc$ radius.
The spiral arms are superimposed on the main disk.
Abrupt velocity changes of up to
$\sim200\kmps$ are observed along the spiral arms in $S$-shaped twists
of isovelocity contours and double velocity-peaked features on the spectra.
The nuclear disk with $\sim600\pc$ radius appears in position-velocity
(PV) diagrams, having an intense concentration of molecular gas.
Its appearance on PV diagrams is indicative of oval motions of
the gas, rather than circular. The nuclear disk and spiral arms
form the so-called ``figure-of-eight'' pattern on a PV diagram.
The nuclear core is more compact than our current resolution ($2\arcsec=
150\pc$), and has a gas mass of $3\times10^8\Msun$ within the central
$150\pc$. Though it is unresolved, the nuclear core shows a very high
velocity $\sim200\kmps$ even at the radius of
$\sim100\pc$ on the PV diagram.

We propose a model that NGC 3079 contains a weak bar.
The weak bar model explains the observed features of the main disk,
spiral arms, and nuclear disk. The main disk and spiral
arms result from gaseous $x_1$-orbits and associated crowding respectively.
The nuclear disk arises from gaseous $x_2$-orbits.
The gas concentration in the nuclear disk could be explained by the
expected gas-fueling mechanism: the gas on $x_1$-orbits flows along
spiral arms (or offset shocks), colliding with the gas on $x_2$-orbits,
and accumulating onto the nuclear disk. Assuming that the gas moves
nearly along the spiral arms which run perpendicular to
the line-of-sight, the pattern speed of the bar is estimated to be
$55\pm10 \kmpspkpc$. The high velocity of the nuclear core cannot be
explained by our model for a bar. Thus we attribute it to a central massive
core with a dynamical mass of
$10^9\Msun$ within the central $100\pc$. This mass is three orders
of magnitude more massive than that of a central black hole
in this galaxy.

\end{abstract}

\keywords{
galaxies: active ---
galaxies: ISM ---
galaxies: kinematics and dynamics ---
galaxies: spiral ---
galaxies: structure
}

\section{INTRODUCTION}

NGC 3079 is a nearly edge-on ($i=77\arcdeg$) SBc galaxy at a distance
of $15.6\mpc$ \citep[1 arcsec corresponds to $76\pc$; ][]{sf99}.

Its nucleus is classified as LINER \citep{hk80} or Seyfert 2
\citep{fd86,ho97,sb01} from the optical emission spectrum, and shows
strong $\rm H_2O$ maser emission \citep{hk84,na95,tr98}.
The overall velocity distribution of $\rm H_2O$ masers suggests
the presence of a binding mass of $\sim10^6\Msun$ at the center
\citep{tr98}, possibly a central supermassive black hole,
which has been found in many AGNs \citep{mi95,wa99,is01}. Parsec-scale
nuclear jets observed in the radio continuum may be outflow from
the central compact object \citep{is88, ss00}.

\begin{figure}
\plotone{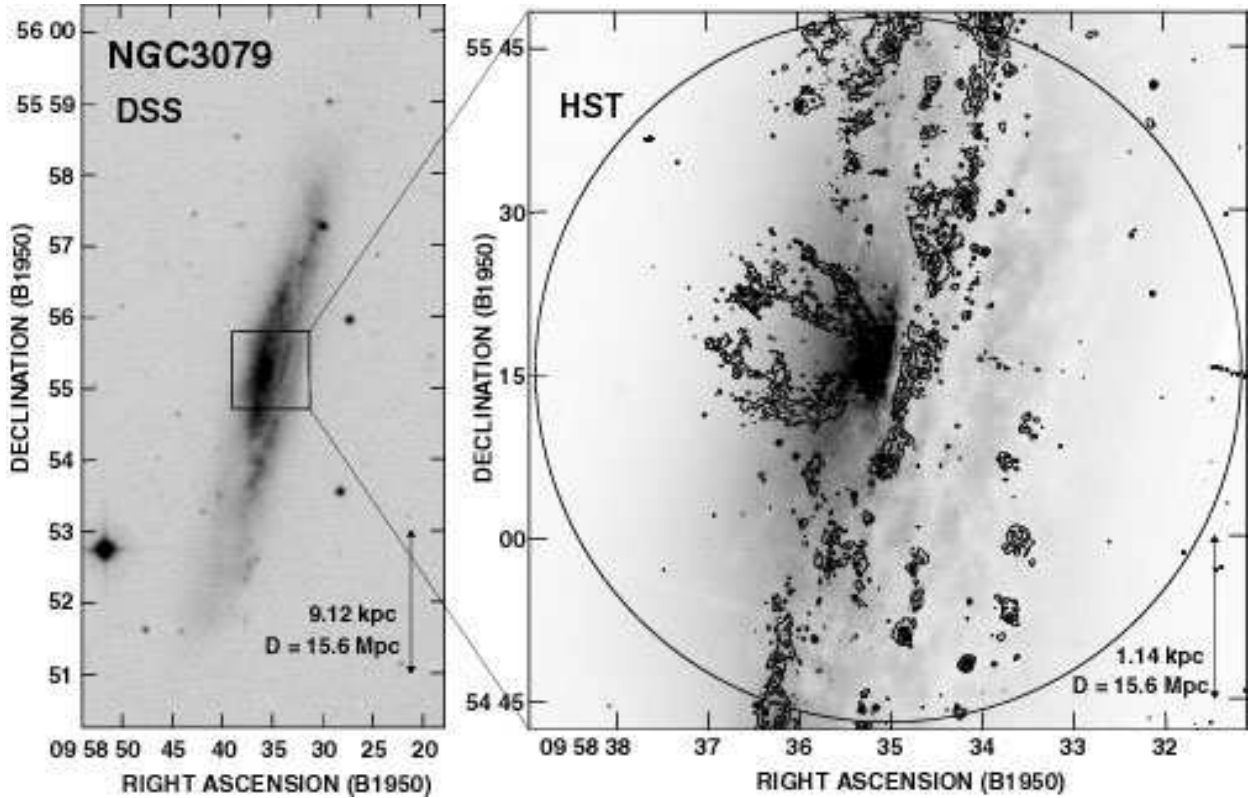}
\caption{Optical images of NGC 3079. (a) R-band image from the Digitalized
Sky Survey. (b) HST WFPC2 images of $I$-band (grey scale) and
$\Halpha+\NII$ line emission (contour), obtained from the HST data archives
(P.I. G. Cecil). Contours are at 2, 5, 15, 20, 30, 40, 50, 60, 70, 80, 90,
100\% of the $\Halpha+\NII$ peak intensity. The circle represents the
primary beam size of our CO (1-0) observations (65'' HPBW).
The absolute positions of both panels were calibrated using the USNO-A2.0
catalog \citep{usno2}, and uncertain to $0\arcsec.5$.
\label{fig:optimgs}}
\end{figure}

\placefigure{fig:optimgs}
Kiloparsec-scale significant outflow along the galaxy minor axis is
observed as lobes in the radio continuum \citep{ds88}, $\Halpha$ emission
\citep[see Figure \ref{fig:optimgs}]{fd86,vl94,ce01}, and X-ray
emission \citep{fb92,pt98}. Optical spectroscopy shows gas
motions with a velocity range of $\sim2000\kmps$ across the lobes,
and unusually high $\NII/\Halpha$ line ratios which indicate the
presence of shocks \citep{vl94}. This type of large-scale outflow
is often attributed to starburst activity \citep{hk90}.
However there are some arguments against this starburst model for
NGC 3079 \citep{hw95}. \citet{ds88} present an alternative model
that a wind flow from the nucleus could be directed toward the
galaxy minor axis by interaction with dense gas surrounding the
nucleus. This interaction produces shocks which explain the observed
strength of $\rm H_2$ emission from lower vibration transitions
by collisional excitation \citep{hw95}.

At the root of the kiloparsec-scale outflow, there is a dense
molecular disk with a radius of a few kiloparsec \citep{ycs88,
si92, is92, sf01}. Based on their interferometry observations
of CO ($J=1-0$) emission,
\citet{si92} found an intense concentration of gas in the
molecular disk, spiral arm features visible on
a position-velocity diagram, and a central compact core
which was unresolved at their resolution ($4\arcsec$). Recently
\citet{sf01} have confirmed these features, and further resolved
and classified the innermost region into the nuclear molecular
disk and ultra-high-density core, based on their appearance
on intensity maps and position-velocity diagrams. [In this paper,
we will re-classify the features based on our new data
(\S \ref{sec:result}) and on theoretical considerations
(\S \ref{sec:dyn}).] The molecular disk was also observed in
HCN and HCO$^+$ \citep{kn01}, which measure the
concentration of high density molecular clouds toward the nucleus.


The disk of NGC 3079 is rotating with the north-side approaching and
south-side receding. A $K^\prime$-band image shows spiral arms
in the disk, forming an inverted {\it S}-shaped pattern on the sky
\citep{vl99}.
If trailing spiral arms are assumed, the west of NGC 3079 is
near-side, which is consistent with the entire dust lane morphology.
NGC 3079 is a far-infrared (FIR) luminous galaxy listed
in the {\it IRAS} bright galaxy catalog \citep{so89}. An {\it ISO}
$90\mu m$ map shows that most of the FIR emission is produced by
dust heated by stars in the entire galaxy disk, rather than the
nucleus \citep{pg00}. The HI disk extends more widely than the
optical disk and shows warps at the outskirts \citep{is91}.
These warps may originate from an interaction with a nearby
companion NGC 3073. The HI velocity field, however, shows
a regular rotating-disk pattern. The radial emission profile
is well fitted by an exponential function, while it shows a sharp
drop in the central region \citep[$<50\arcsec$; ][]{is91}
which is produced by absorption against strong continuum emission
\citep{ds88}. The CO disk is embedded in this region of the absorbed
HI \citep{ycs88, si92, sf99}, and coincident with a void of $\Halpha$
line emission \citep{ce01}.

Though NGC 3079 is classified as a barred galaxy (SBc), the stellar
bar is hardly confirmed in optical/infrared photographs
in this nearly edge-on galaxy. There is however some evidence for
the presence of a bar. NGC 3079 has a stellar bulge which takes
a so-called peanut-shape \citep{sw93}, whose possible origin is
a vertical instability of rotating disk-stars in a bar potential
\citep{cm90}. \citet{vl99} fitted an oval orbit model to their $\Halpha$
velocity field, and concluded that bar streaming motions with moderately
eccentric orbits ($e=b/a\sim0.7$) aligned along ${\rm P.A.}=
130\arcdeg$ intrinsic to the disk (${\rm P.A.}=163\arcdeg$ on the sky
\footnote{
The bar position angle on the sky ($\rm P.A._{sky}$) is calculated from
that intrinsic to the disk ($\rm P.A._{int}$), by
``$\tan({\rm P.A._{gal}- P.A._{sky}}) = \tan({\rm P.A._{gal}-P.A._{int}})
\cos i$'' using their adopted position angle and inclination of the disk.})
are satisfactory to match the observations.

\placetable{tab:prop}

In this paper, we present our recent $^{12}$CO ($J=1-0$) observations of the
central $1\arcmin$ of NGC 3079 using the Nobeyama Millimeter Array (NMA).
We describe the observations and data analysis in \S \ref{sec:nmaobs}.
A part of the data has been published in \citet{sf01}, but this paper
presents our new analyses including new observations.
The main features in the molecular disk of NGC 3079 are presented
in \S \ref{sec:result}. Gas dynamics in the molecular disk are discussed
in \S \ref{sec:dyn}. A weak bar and central massive core explain all
the observed features. The central rotation curve and mass are
derived in \S \ref{sec:dms}. We summarize our conclusions in
\S \ref{sec:conclusion}.

\section{OBSERVATIONAL DATA}\label{sec:nmaobs}

\subsection{NMA $^{12}$CO ($J=1-0$) Observations}

Our aperture synthesis observations of the $^{12}$CO ($J=1-0$) emission from
NGC 3079 were obtained with the Nobeyama Millimeter Array (NMA) at
the Nobeyama Radio Observatory (NRO)
\footnote{Nobeyama Radio Observatory is a branch of the National
Astronomical Observatory, operated by the Ministry of Education,
Culture, Sports, Science and Technology, Japan.},
between 2000 January and 2001 April for a single pointing center at
($\alpha_{1950}$, $\delta_{1950}$) = ($9^{\rm h}58^{\rm m}35^{\rm s}.02$,
$+55\arcdeg55\arcmin15\arcsec.40$).
We made the observations with three available configurations (AB, C, and D);
when combined, the visibility data cover projected baselines from 10
to 351 m. The NMA consists of six 10 m antennas, providing a FWHP
of about $65''$ at 115 GHz. The antenna size limits the minimum projected
baseline, restricting the largest detectable size of objects to about
$54\arcsec$. Tunerless SIS receivers have receiver noise temperatures
of about $30\K$ in double sideband, and typical system noise
temperatures of about $400\K$ in single sideband.
Digital spectro-correlators \citep{ok00} have two spectroscopic modes;
we used a mode covering 512 MHz ($1331\kmps$) with
2 MHz ($5.2\kmps$) resolution. We observed the quasar 0954+556 every 20
minutes for gain calibration,
and 3C279 (or 3C273) for bandpass calibration. Absolute flux scales
(0.68 Jy at 115 GHz for 0954+556, uncertain to $\sim 20\%$) were
measured three times in 2000 and once in 2001; no significant flux
variation was observed.

The raw visibility data were calibrated for complex gain and passband
with the UVPROC-II package developed at NRO, and mapped with
the NRAO/AIPS package. We applied the CLEAN procedure with natural
weighting for each velocity channel, and obtained a three-dimensional
(RA, DEC, $\vlsr$) data cube. Parameters of the cube are listed
in Table \ref{tab:cube}. 

\placetable{tab:cube}

\begin{figure}
\epsscale{0.6}
\plotone{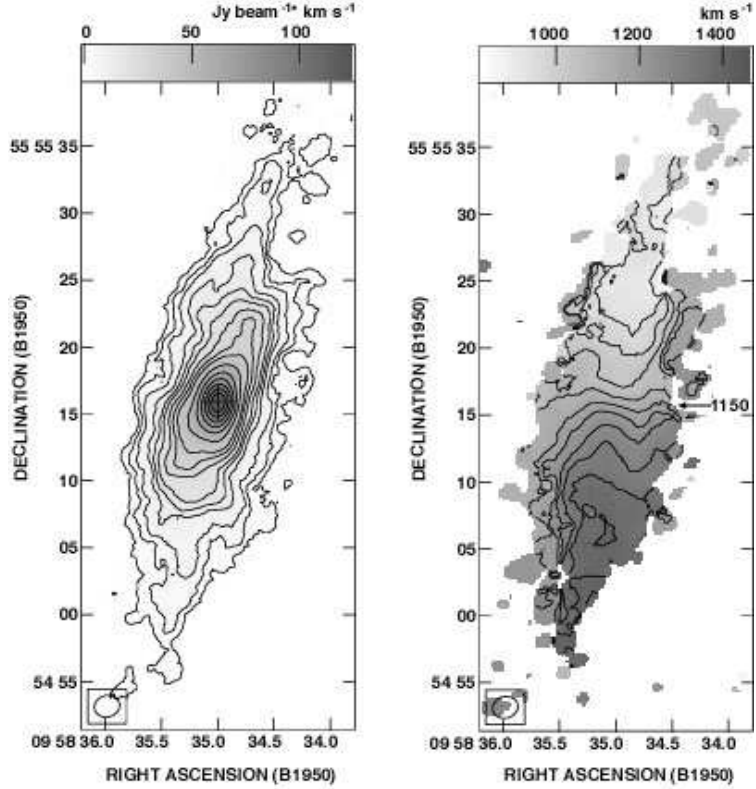}
\caption{CO (1-0) zeroth- and first-moment maps in the central $20\arcsec
\times 48\arcsec$ ($1.5\kpc \times 3.7\kpc$) region of NGC 3079.
The synthesized beam ($1.9\arcsec \times 1.6\arcsec$)
is shown in the lower left corners.
{\it Left:} The zeroth-moment map (integrated-intensity map).
Contour levels are 1.5, 3, 5, 9, 12, 15, 20, 25, 30, 40, 50, 60, 70, 80, 90,
100 \% of the peak integrated-intensity $1.25 \times 10^2 \jypbm \, \kmps$.
The clip level was set to $2.5\sigma$ where $1 \sigma = 12 \mjypbm$.
The primary beam attenuation has not been corrected.
{\it Right:} The first-moment map (intensity-weighted mean velocity map).
Contours are drawn with an interval of $50\kmps$; The contour of $1150\kmps$
($\vsys= 1147\kmps$) is pointed to by an arrow. The clip level was set to
$4\sigma$ to make this map.
\label{fig:mom01}}
\end{figure}

Figure \ref{fig:mom01} displays zeroth- and first-moment maps in the
central $20''\times 48''$ region ($1.5\kpc \times 3.7\kpc$) of
NGC 3079, while Figure \ref{fig:chmap} shows channel maps with an
interval of $10.4\kmps$ in the same region. The significant emission
($>3\sigma$; $1\sigma=12\mjypbm$) is detected in 63 adjacent channels
within the velocity range of $\vlsr=821-1467\kmps$ ($\Delta V = 646\kmps$).
No primary beam correction has been applied in these maps.
Figure \ref{fig:pvd} displays a position-velocity diagram (PV diagram)
along the optically defined major axis (P.A.$=165\arcdeg$), integrated
along the minor axis. The almost entire CO emission in Figure
\ref{fig:mom01} ({\it left}) falls in the slit-width ($12\arcsec$).
The axes are labeled relative to the dynamical center,
and systemic recession velocity of the galaxy
(derived in $\S$ \ref{sec:diskkin}; see Table \ref{tab:kin}).

\placefigure{fig:mom01}
\placefigure{fig:chmap}
\placefigure{fig:pvd}

Compared with single-dish observations from IRAM 30 m and FCRAO 14 m
telescopes \citep{br93,yg95}, our NMA cube recovers about 87\% of
the total CO line flux in the central $23''$ and 67\% in $45''$,
which means that the CO emission is concentrated in the central region.
We made no correction for the missing flux in the following discussion.
Figure \ref{fig:spec} compares a spectrum from our NMA cube with that from
the IRAM observations \citep{br93}.
In order to obtain the fluxes and spectrum comparable to the single-dish
ones, the NMA cube was corrected for the primary beam response of
antennas, convolved with the single-dish beam ($23''$ for IRAM and
$45''$ for FCRAO), and sampled at the pointing center of the single-dish
observations.

\placefigure{fig:spec}

\subsection{Supplied HST Data}
We obtained images from the {\it Hubble Space Telescope (HST)} archive
\citep[P.I. G. Cecil; see ][]{ce01}. The images were taken with WFPC2 and
two filters, F814W and F658N, which correspond to the $I$-band and
$\Halpha+\NII$ filters in ground telescopes. The central part of NGC 3079
lies on a WFC chip rather than the PC chip.
Cosmic-ray hits were removed and images are combined with the IRAF/STSDAS
package. The absolute position of the $HST$ images is calibrated using
the USNO-A2.0 catalog \citep{usno2}, and is accurate to about $0''.5$.
The derived images are presented in the right-hand panel of Figure
\ref{fig:optimgs}.

\section{RESULTS}\label{sec:result}

Our maps and PV diagram show four distinct components: a main disk,
spiral arms (or offset ridges), a nuclear disk, and a nuclear core.
Figure \ref{fig:schpvd} shows a schematic illustration of these four
components on a PV diagram. We describe each of the four in order of
decreasing radius in \S \ref{sec:mdsk}-\ref{sec:core}, and compare
them with previous results in \S \ref{sec:cmpprv}.

\placefigure{fig:schpvd}

\subsection{The Main Disk}\label{sec:mdsk}

\subsubsection{Smooth Gas Distribution on the Main Disk}\label{sec:diskdis}

The zeroth-moment map of Figure \ref{fig:mom01} displays a disk with
an extent of $45\arcsec \times 15\arcsec$ ($3.4\kpc \times 1.1\kpc$),
elongated along the optically-defined major axis of the galaxy
($\mbox{P.A.} = 165\arcdeg$). Since our synthesis observations have
an intrinsic maximum limit of detectable scale ($54\arcsec$),
larger components detected in single-dish observations
\footnote{Recently, NRO 45 m telescope observations found that the
full extent of CO emission in NGC 3079 is about $80\arcsec$ in radius
(Yamauchi, private communication).}
are not covered in this map, and result in the missing flux (our
total flux recovery is about 67\% in $45\arcsec$).
The inclination of the galaxy ($77\arcdeg$) is large, but enough
to show the emission distribution on the disk, because the vertical sizes
of molecular disks are usually thin ($150\pc$ for the Galaxy from
\citealt{css88}; $230\pc$ for NGC 891 from \citealt{hn92}) in comparison
with the full extents of molecular
disks ($3.4\kpc$ or $45\arcsec$ for NGC 3079 from the zeroth-moment map).
The CO emission is smoothly distributed on the full disk, showing no
emission deficient regions often found in molecular disks \citep[see ][]
{sk99, re01}.
The emission distribution is nearly axisymmetric, except for slight
enhancements owing to spiral arms (see $\S$ \ref{sec:arms}).
The main disk coincides with the void of HII regions \citep[see also
Figure \ref{fig:haco} {\it left}]{ce01}, and with the region where
the HI gas is observed in absorption \citep{is91} against strong radio
continuum emission \citep{ds88}. The center of the main disk coincides
with the root of the $\Halpha$-lobe.

\placefigure{fig:haco}

\subsubsection{Kinematics of The Main Disk}\label{sec:diskkin}

The first-moment map (Figure \ref{fig:mom01} {\it right}) shows a rotating
disk with the northside approaching and southside receding.
The $S$-shaped twists of isovelocity contours should be attributed to
spiral arms ($\S$ \ref{sec:arms}).
The entire velocity field is almost perfectly bisymmetric, indicating
regular motions in the molecular disk.

We obtained kinematical parameters from the first-moment map using the
AIPS/GAL package. The dynamical center, position angle and inclination
are first determined using the Brandt rotation curve model \citep{br65}.
Then we fixed the above parameters, fitted tilted ring models with
constant velocities to the first-moment map, and obtained the systemic
recession velocity (LSR velocity). The results are listed in Table
\ref{tab:kin}.
The dynamical center coincides spatially with the emission centroid:
($\alpha_{1950}$, $\delta_{1950}$) = ($09^{\rm h}58^{\rm m}35^{\rm s}.00$,
$+55^{\rm d}55^{\rm m}15^{\rm s}.80$).
The position angle and inclination are in good agreement with the results
from optical isophotes \citep{st81, vl99} and from kinematics of the HI
gas \citep{is91}. The systemic recession velocity is consistent with
estimates from optical spectroscopic data
\citep[e.g. $1150\kmps$ from][]{vl99},
but slightly exceeds the value from $\HI$ data
\citep[$1124\kmps$ from][]{is91}. This difference might come from a warp
of the HI disk \citep{is91}. We adopt the values from large-scale
isophotes for position angle and inclination, and the derived values
for galaxy center and systemic recession velocity in the rest of
our analysis. These choices do not affect the following discussion.

\placetable{tab:kin}

The PV diagram (Figure \ref{fig:pvd}) shows the main disk as two ridges,
symmetrically extending southeast and northwest from the velocity peaks
around the center (see also Figure \ref{fig:schpvd}). The near-perfect
symmetry in the diagram again indicates regular motions of the molecular
gas in the disk. The velocity of $220\kmps$ at the radius of $26\arcsec$
($2.0\kpc$) indicates the dynamical mass of $M_{\rm dyn}=2.2 \times 10^{10}
\Msun$. The disk is also confirmed in the channel maps
(Figure \ref{fig:chmap}); the approaching (NW) side appears in the channels
of $\vlsr = 821-1029\kmps$, while the receding (SE) side is found in
$\vlsr = 1258-1467\kmps$.

\subsubsection{Mass and Surface Densities of the Molecular Gas}

Masses of molecular gas $M_{\rm H_2}$ are estimated from total CO-line
flux $S_{\rm CO}$, CO-to-${\rm H}_2$ conversion factor $X_{\rm CO}$,
and galaxy distance $D$ by
\begin{eqnarray}
\left( \frac{M_{\rm H_{2}}}{\Msun} \right)
&=& 7.2 \times 10^3
\left( \frac{D}{\mpc} \right)^2
\left( \frac{S_{\rm CO}}{\jykmps} \right) \nonumber \\
& & \times \left( \frac{X_{\rm CO}}{1.8 \times 10^{20} \unitcnv} \right).
\end{eqnarray}
Assuming the hydrogen mass fraction \citep[0.707 from][]{al00},
the total gas mass, including He and other elements, becomes
$M_{\rm gas} = 1.41 M_{\rm H_2}$.
Face-on surface densities of molecular gas are calculated from
the integrated CO line intensity $I_{\rm CO}\Delta V$,
galaxy inclination $i$, and conversion factor $X_{\rm CO}$ by
\begin{eqnarray}
\left( \frac{\Sigma_{{\rm H}_2}}{\Msun \, {\rm pc^{-2}}} \right)
&=& 3.0 \times 10^2 \cos i
\left( \frac{I_{\rm CO} \Delta V}{\jykmps \, {\rm arcsec^{-2}}} \right) \nonumber\\
& &\times \left( \frac{X_{\rm CO}}{1.8 \times 10^{20} \unitcnv} \right).
\label{eq:srfd}
\end{eqnarray}
The surface density of total gas also becomes
$\Sigma_{\rm gas} = 1.41 \Sigma_{\rm H_2}$.
The surface density estimation is independent of galaxy distance.
We adopt the conversion factor of $X_{\rm CO} = 1.8 \times 10^{20} \unitcnv$
from observations in the Galaxy \citep{dm01}, while $X_{\rm CO}$ could be
smaller in galactic center regions \citep{ar96}.

The total flux within the radius of $26\arcsec$ is $S_{\rm CO}= 1.9 \times
10^3 \jykmps$, which corresponds to $M_{\rm gas} = 4.6\times 10^9\Msun$
for the galaxy distance $D=15.6\mpc$. The ratio of dynamical mass
($2.2\times 10^{10}\Msun$ from \S \ref{sec:diskkin}) to the total
molecular gas mass
$M_{\rm gas}/M_{\rm dyn}$ is about 21\% within the radius of $2\kpc$.
The peak integrated intensity is
$I_{\rm CO}\Delta V = 1.3 \times 10^2 \jypbm \, \kmps$
(or $3.8 \times 10^3 \kkmps$) at ($\alpha_{1950}$, $\delta_{1950}$)
= ($09^{\rm h}58^{\rm m}35^{\rm s}.00$, $+55\arcdeg55\arcmin15\arcsec.80$);
the corresponding face-on surface density becomes $\Sigma_{\rm gas} = 2.5
\times 10^3\Msun \pc^{-2}$. No correction for missing flux is applied.

Figure \ref{fig:srfd} shows the radial profile of the CO emission. Integrated
intensities are azimuthally averaged in each annulus ($\Delta r=0.5\arcsec$)
with corrections for inclination ($77\arcdeg$) and primary beam attenuation.
The outskirts of the profile ($r>5\arcsec$) is well fitted by an exponential
(dotted line). At the innermost region, the CO emission exceeds this exponential
curve, indicating a high CO concentration at the central region.
Surface densities of molecular gas are labeled at the right-hand axis.
The observed value of $\Sigma_{\rm H_2} = 1000 {\rm\,\Msun pc^{-2}}$
at $r \sim 5\arcsec$ ($380\pc$) is an order of magnitude higher than
the typical value for the central regions of nearby galaxies
\citep{sk99b}.

\placefigure{fig:srfd}

The $\HI$ gas surface density also follows an exponential profile
$\Sigma_{\rm HI}^0 \, e^{-r/h_{\rm HI}}$ where $h_{\rm HI}=118\arcsec$
($9.0\kpc$) and $\Sigma_{\rm HI}^0 = 8.8 \times 10 {\,\Msun \pc^{-2}}$
\citep[from their Table 4]{is91}. In comparison with this HI profile,
our ${\rm H_2}$ profile (dotted line) takes the same surface density at
nearly the edge of our detected molecular disk ($r\sim23\arcsec$ or
$1.7\kpc$), rising more steeply toward the galaxy center
($h_{\rm CO}\sim7.2\arcsec$ or $550\pc$).
This indicates that the gas in NGC 3079 is highly concentrated in
the central molecular disk.

\subsection{The Spiral Arms/Offset Ridges}\label{sec:arms}

The zeroth-moment map (Figure \ref{fig:mom01} {\it left}) shows
arm-like enhancements of emission, superimposed on the main disk;
two ridges bisymmetrically run northwest and southeast from the galaxy
center (most obvious within the central $\sim10\arcsec$ radius),
forming an {\it inverted} $S$-shaped pattern. These features resemble
(trailing) spiral arms, or offset ridges often found at the leading
edge of stellar bars. [Note though these enhancements do arise from
the spiral arms, they are slightly offset from the actual spiral arms
(see Figure \ref{fig:haco}), because this projected map is contaminated
by emission from other components.] The $S$-shaped twists of isovelocity
contours on the first-moment map (Figure \ref{fig:mom01} {\it left})
coincide with the two arms, indicating non-circular motions around
the spiral arms. Similar inverted $S$-shaped spiral arms have been
found in a $K'$-band image \citep{vl99}.

Figure \ref{fig:lineprof} shows line-profiles of CO emission at $5
\times 7$ points on a $3\arcsec$-spacing grid (crosses on the left-hand
panel), centered at the dynamical center (Table \ref{tab:kin}).
Remarkable double-peaked features appear along the spiral arms:
arrows point to examples of the double-peaks.
The northwest arm corresponds to peaks with higher velocities,
while the southeast arm corresponds to peaks with lower velocities.
The other peaks come from the main disk. These distinct velocities
between the main disk and spiral arms (typically $\Delta V\sim200\kmps$)
produce the $S$-shaped twists in the first-moment map.
Similar double-peaked features have been noticed from
$\Halpha$-spectroscopic data \citep{fs92, vl99}, though their low velocity
resolution did not allow them to distinguish between the main disk
and spiral arms.

\placefigure{fig:lineprof}

The spiral arms are also confirmed on the PV diagram (Figure \ref{fig:pvd};
see also Figure \ref{fig:schpvd}). Two ridges run northwest and southeast
with lower velocities than those of the main disk. These two ridges and
a central ridge of the nuclear disk (\S \ref{sec:nstrm}) show a so-called
``figure-of-eight'' pattern (or ``tilted X'' pattern). Similar patterns
are often found in edge-on barred galaxies \citep{hn90, ln90, gb95, mk99}.
In the velocity range of $\Delta V = \pm 150\kmps$ ($\vsys=1147\kmps$),
the spiral arms are more dominant in emission than the main disk.

This is also evident in the channel maps (Figure \ref{fig:chmap}).
The spiral arms are confirmed at $V = 1010-1280\kmps$; emission
from the northwest and southeast arms appear at the northwest and
southeast edges of maps at $1010$ and $1280\kmps$, approaching
the innermost core with increasing and decreasing velocities, respectively.
Figure \ref{fig:haco} (right-hand panel) shows contours of a zeroth-moment
map made in the velocity range of $\Delta V = \pm 100\kmps$ centered at
the systemic recession velocity (superimposed on the $\Halpha+\NII$ map).
The spiral arms are rooted to the nucleus, forming an inverted $S$-shaped
pattern on the sky.

\subsection{The Nuclear Disk}\label{sec:nstrm}

The nuclear disk shows a sign of an existence on the PV diagram
(Figure \ref{fig:pvd}): remarkable velocity-peaks appear 
symmetrically at ($-7\arcsec$, $+250\kmps$) and ($+7\arcsec$,
$-250\kmps$). Figure \ref{fig:ewpvd} shows PV diagrams with
$1\arcsec$-width and offsets of $+2\arcsec$, $0\arcsec$ and
$-2\arcsec$ along the galaxy minor axis. The cuts are along
the galaxy major axis. These diagrams reveal that the two
peaks are on a single kinematical sequence: ridges are running
and connecting the two peaks through three panels (see also
Figure \ref{fig:schpvd}).
This sequence seems to be rotating in this innermost region
and forming a nuclear disk. The nuclear disk is obviously
distinct from other velocity-peaks in the innermost region
(at $\pm1\arcsec$ in Figure \ref{fig:pvd}), because their
corresponding peaks appear only in the diagram with
$0\arcsec$-offset; we thus think that they are consequences of
the nuclear core (see $\S$ \ref{sec:core}). The very high CO
concentration within the nuclear disk ($r<7\arcsec$; Figure
\ref{fig:pvd} and \ref{fig:srfd}) indicates its distinct
nature from the main disk. We conclude that the component which
appears as the peaks and ridges in the PV diagrams is the nuclear
disk. This component is rotating around the galaxy center with
the northside approaching and southside receding.

\placefigure{fig:ewpvd}

In the three panels of Figure \ref{fig:ewpvd}, the ridges corresponding
to the nuclear disk are not on a straight line connecting their two
velocity peaks:
the ridges are bent downward and upward in the $+2\arcsec$
and $-2\arcsec$ offset diagrams respectively, while the ridge twists
upward (at the left-hand side) and downward (right-hand side)
in the $0\arcsec$-offset diagram.
The deviations from the straight line amount up to
$\sim 50\kmps$. These symmetrical bends and twist indicate
that the gas in the nuclear disk takes oval orbits rather than
circular (see $\S$ \ref{sec:dyn}).

The nuclear disk is also seen in the channel maps (Figure \ref{fig:chmap}).
The two velocity-peaks on the PV diagrams appears in the channels of
$\vlsr = 822-926\kmps$ as an emission peak at around $7\arcsec$-north
from the center, and in $\vlsr = 1384-1467\kmps$ at around $7\arcsec$-south.
In other channels, strong emission from the main disk
and nuclear core contaminates the nuclear disk emission.

\subsection{The Nuclear Core}\label{sec:core}

The nuclear core is most clearly seen in the PV diagram (Figure
\ref{fig:pvd}), which appears as two velocity-peaks at ($-1\arcsec$,
$+240\kmps$) and ($+1\arcsec$, $-240\kmps$).
The distinct kinematics of this core from the other components
is evident in a series of PV diagrams which have positional
offsets along the galaxy minor axis (Figure \ref{fig:ewpvd}).
Among the three diagrams with $+2\arcsec$, $0\arcsec$ and
$-2\arcsec$-offsets, the nuclear core appears only in the central
one. The high velocities indicate a large dynamical mass in this nuclear
region if the gas is pure-circularly rotating; we will discuss this
in $\S$ \ref{sec:orgcore}.

The zeroth-moment map (Figure \ref{fig:mom01}) shows the nuclear core
as an intense CO concentration at the very center of the galaxy.
It is also confirmed in channel maps (Figure \ref{fig:chmap}) in
almost all the velocities ($\vlsr=863-1436\kmps$).
The central gas density of the nuclear core is as high as
$2.5\times 10^3 \Msun \, \pc^{-2}$, which is about an order of
magnitude higher than is typical for nearby normal galaxies
\citep{sk99}.
This value is however a lower limit, because the core is unresolved
even in our high resolution observations ($1.9\arcsec \times 1.6\arcsec$),
The total flux within the radius of $2\arcsec$ ($152\pc$) is
$S_{\rm CO} = 1.2 \times 10^2 \jykmps$; the corresponding gas mass
is $M_{\rm gas} = 3.0 \times 10^8 \Msun$.

\subsection{Comparison with Previous Results}\label{sec:cmpprv}


On the molecular disk of NGC 3079, \citet{si92} identified three main
components: an outer disk, spiral arms, and a central compact core.
Recently \citet{sf01} further resolved the central compact core,
and classified it into two distinct components, a nuclear molecular disk
and ultra-high-density molecular core, based on their appearance
on maps and PV diagrams. In this paper, we re-classified the observed
features into four distinct components, based on our new observations and
analyses, and partly on theoretical considerations in \S \ref{sec:dyn}.
The outer disk and spiral arms in \citet{si92} correspond to our main
disk and spiral arms respectively, and the ultra-high-density molecular
core in \citet{sf01} corresponds to our nuclear core. The nuclear molecular
disk in \citet{sf01} is different from our nuclear disk, and includes
both our nuclear disk and main disk in part.


We confirmed two peaks on a PV diagram in the central region, reported by
\citet{si92}, in Figure \ref{fig:ewpvd}. This feature however cannot
be confidently attributed to the nuclear ring \citep{si92} because
the two peaks may arise from our nuclear disk (see Figure \ref{fig:ewpvd}
{\it top} and {\it bottom}) superposed on our nuclear core.
Our central surface density $\Sigma_{\rm gas}=2.5\times10^3\Msun\,\pc^{-2}$
is consistent with the value ($2.2\times10^4\Msun\,\pc^{-2}$) of \citet{si92}
by considering the difference of the adopted conversion factors and
applying the correction for inclination to the value of \citet{si92}.
We could not find unambiguous evidence for molecular outflow \citep{is92}
in our data.

\section{GAS DYNAMICS IN NGC 3079}\label{sec:dyn}

The near-perfect bisymmetric structures of the molecular disk are striking.
These kinds of bisymmetry are often thought to be consequences of gas
motions in a bar. Though a stellar bar is not seen in optical/infrared
photographs in this nearly edge-on galaxy, there is some indirect
evidence for the presence of a bar: the peanut-shaped
bulge \citep{cm90, sw93}; the ``figure-of-eight'' pattern on the
PV diagram \citep{km95}; the high central concentration of
the molecular gas \citep{sk99b}; the $\Halpha$ velocity field
well-fitted by an oval-orbit model in a weak bar \citep{vl99}.
We thus propose a model that NGC 3079 contains a weak bar.
This model naturally explains the features in $\S$ \ref{sec:result}:
the main disk and spiral arms (or offset ridges) may result from
gaseous $x_1$-orbits and associated crowding respectively. The nuclear
disk may arise from gaseous $x_2$-orbits. This model however cannot
interpret the nuclear core, which necessitates a massive component
besides the bar at the galaxy center.
Figure \ref{fig:schview} shows schematic views of
our interpretation, drawn on our model orbits (see a description
in \S \ref{sec:dom}). We present below our detailed case
for the existence of the weak bar, and discuss the origin of the main
disk, spiral arms, nuclear disk and core.

\placefigure{fig:schview}

\subsection{Damped Orbit Model}\label{sec:dom}

We adopt a damped orbit model by \citet{wa94} and \citet{sk00}
to describe gas motions in a weak bar potential. This model solves
the equation of motion which includes a damping-force term to emulate
the collisional nature of the gas, and obtains closed orbits for
gas in a bar. This model has been applied to a galaxy NGC 5005 \citep{sk00},
and similar models have been successfully
used to infer the presence of bars in edge-on galaxies which show
characteristic patterns on PV diagrams \citep{bn91, al95, gb95, km95}.

Figure \ref{fig:orbits} displays our model of gas orbits in a bar potential
that reproduces the main features in the zeroth-, first-moment maps, and PV
diagram of NGC 3079. We here took the same potential model as \citet{sk00}
of
\begin{equation}
\Phi(r, \theta) = (1-\epsilon \cos 2 \theta) \frac{v_0^2}{2} \log(1+(r/a)^2),
\end{equation}
where the first and second terms stand for disk and bar potentials
respectively, and $\theta$ is the azimuthal angle from the bar major axis.
For this particular case we adopted the characteristic
radius $a=6\arcsec$ (456\pc), rotation velocity $v_0=240\kmps$ and bar
strength $\epsilon = 0.04$ to match the observations. We chose the pattern
speed of $\omgb=55\kmpspkpc$ based on consideration given in
$\S$ \ref{sec:orgx1}. The inner and outer ILRs and corotation occur
at the radii of $7.7\arcsec$ ($582\pc$), $10.5\arcsec$ ($798\pc$) and
$57.1\arcsec$ ($4.34\kpc$) respectively. We assumed that the position
angle of the bar (dashed line) from the north is ${\rm P.A._{int}} =
135\arcdeg$ intrinsic to the disk, and ${\rm P.A._{sky}} = 158\arcdeg$
on the sky. On the galaxy disk, this position angle of the bar differs
by $30\arcdeg$ from that of the galaxy major axis which runs
horizontally in Figure \ref{fig:orbits} ({\it top}).

\placefigure{fig:orbits}

Our parameters of the bar chosen to match the CO observations are in
good agreement with the values that \citet{vl99} have derived by fitting
an oval-orbit model to their $\Halpha$ velocity field of NGC 3079
($\omgb\sim60\kmpspkpc$;  ${\rm P.A._{bar}}=130\pm10\arcdeg$;
axes ratio of oval orbits 0.7).
Beyond our particular choices of the parameters, the features of gaseous
closed orbits described below are general \citep{wa94,ll94}, and have been
established in numerical simulations \citep{wa94, wk01, kw02}.

Closed orbits in a bar have an elongation in general; the position angle of
the elongation $\psi$ is measured clockwise from the bar major axis
(dashed line in Figure \ref{fig:orbits} {\it top}), in the rest of
the discussion. For collisionless star-like particles, two families of
closed orbits are dominant: $x_1$-orbits have an elongation with
$\psi=0\arcdeg$
and dominate outside of the outer ILR and inside of the inner ILR,
while $x_2$-orbits elongate with $\psi=90\arcdeg$ and are dominant
between the ILRs. Orbits for gaseous particles show similar families
at the same loci (Figure \ref{fig:orbits} {\it top}), while gaseous
$x_1$-orbits have an elongation toward the leading edge of the bar
($\psi>0\arcdeg$), and gaseous $x_2$-orbits elongate opposite
($\psi<90\arcdeg$), owing to the damping force.

\subsection{The Origin of The Main Disk and Spiral Arms}\label{sec:orgx1}


The main disk and spiral arms may originate from gaseous $x_1$-orbits.
The position angle $\psi$ of orbits changes with radius owing to
the damping force, producing crowded regions of the orbits
(Figure \ref{fig:orbits}).
The crowding of the gaseous $x_1$-orbits which appear at the leading edge
of the bar result in the spiral arms (or so-called offset ridges),
which are similar to those on the molecular disk of NGC 3079.
The slightly enhanced CO emission in the zeroth-moment map of
NGC 3079 (Figure \ref{fig:mom01}) may arise from the high gas densities
in the crowded regions. The streamlines belonging to the $x_1$-orbits
extend entirely to the disk, producing the main disk.
 

Our model streamlines have sharp turns at the spiral arms.
Such features, accompanied by shocks, have been observed by
other authors \citep{at92, ab99} in full hydrodynamical simulations.
These turns can produce double-peaked features in line-profiles
observed along the spiral arms (Figure \ref{fig:lineprof}), and
agree spatially with the $S$-shaped twists on the
first-moment map of NGC 3079 (Figure \ref{fig:mom01}).
Figure \ref{fig:pvcomp} shows a comparison between the observed and
model PV diagrams; contours are from the observed one,
while the symbols show the model one, where different symbols
are used for gaseous particles on the spiral arms (squares)
and on the main disk (dots).
On this diagram again, the main disk and spiral arms of NGC 3079
agree well with the gaseous $x_1$-orbits and their crowding respectively.
This match between the observations and the model suggests
that the main disk and spiral arms result from gaseous $x_1$-orbits
and their associated crowding, respectively.

\placefigure{fig:pvcomp}

In the context of our model, gas streamlines on the downstream side
of the sharp turns are nearly perpendicular to our line-of-sight.
Thus the line-of-sight velocity of the gas indicates the pattern
speed of the bar. We estimate the pattern speed from the gradient of
the spiral arms on the observed PV diagram, which appears to be
$\omgb = 55\pm10\kmpspkpc$; this is our adopted value to describe
gas orbits, and is consistent with the result of other authors
\citep[$60\kmpspkpc$ from ][where the coincidence of corotation
and the bar-end was assumed]{vl99}.

\subsection{The Origin of The Nuclear Disk}\label{sec:orgndsk}


The nuclear disk may result from gaseous $x_2$-orbits.
In Figure \ref{fig:pvcomp}, the model PV diagram shows that the gaseous
$x_2$-orbits (crosses) are in good agreement with the nuclear disk of
NGC 3079. Figure \ref{fig:ewmdlpvd} displays PV diagrams along the
galaxy major axis for the orbits within the outer ILR.
The three panels show PV cuts with $10\arcsec$-width (intrinsic to the
galaxy) and positional offsets for the east, central and west sides
of the model galaxy respectively.
These panels reveal that the bends and twist of the ridges discussed
in \S \ref{sec:nstrm} (see Figure \ref{fig:ewpvd}) are also reproduced
in our model:
the top and bottom panels show bends downward and upward respectively,
while the middle panel shows a twist upward (at the left-hand side)
and downward (right-hand). Since a purely rotating disk cannot produce
this type of symmetric bends and twist, the gas on the nuclear disk
must take oval orbits. Our gaseous $x_2$-orbits naturally reproduce
the observed properties of the nuclear disk.

\placefigure{fig:ewmdlpvd}

For later discussion in $\S$ \ref{sec:orgcore}, we mention the position
angle of the $x_2$-orbits on the nuclear disk here in advance. The major
axis of the orbits are inclined counterclockwise to the south from our
line-of-sight in our current configuration (Figure \ref{fig:schview}).
On the other hand, orbits inclined in the opposite direction can also
reproduce the nuclear disk on the PV diagram (Figure \ref{fig:pvd}).
These orbits however make the ridges on the eastern and western PV
diagrams (Figure \ref{fig:ewmdlpvd}) bend in the opposite directions,
and thus cannot match the observations (Figure \ref{fig:ewpvd}).
Therefore the orbits on the nuclear disk must have an inclination
southward (counterclockwise) on the sky from our line-of-sight,
like our gaseous $x_2$-orbits.

Many barred galaxies show an intense concentration of CO emission in
their central parts \citep{sk99b, sk00}. NGC 3079 also has such
concentration in the nuclear disk region (see Figure \ref{fig:srfd}).
This feature is well explained in the scenario of gas motions in bar:
gas flow on $x_1$-orbits like ours leads to offset shocks
at the leading edge of a bar, and results in an inflow of gas toward
the nuclear region. Then the inflowing gas on the $x_1$-orbits collides
with gaseous $x_2$-orbits near the pericenter of the $x_1$-orbits, entering
onto the gaseous $x_2$-orbits, thus increasing mass densities in the nuclear
disk. This scenario has been confirmed in hydrodynamical simulations
\citep{pst95,ab99}.

\subsection{The Origin of The Nuclear Core}\label{sec:orgcore}
Our current model naturally explains the main features of the molecular
disk of NGC 3079, which are the consequences of
gas motions in a weak bar. However the model does not explain the nuclear
core which appears in the PV diagram as high velocity-peaks near
the galaxy center (Figure \ref{fig:pvd}). Owing to our ability to match
the observations with our model for a bar, we attribute the high
velocities to another origin, a massive central component,
besides the bar. There still is a small possibility that
the high velocities may arise from gaseous $x_1$-orbits within the inner
ILR which could elongate along our line-of-sight. However
the gaseous $x_1$-orbits are not the likely origin for two reasons.

(1) Our gaseous $x_1$-orbits within the inner ILR ($7.7\arcsec$) look
near circular, thus cannot show high velocities at their pericenter.
Numerical experiments have also shown similar features: circular orbits
in the innermost regions \citep{pst95, ab99}. In the intensive study of PV
diagrams for galaxies with bars by \citet{ba99} and \citet{ab99},
the $x_1$-orbits within the inner ILR do not produce any remarkable
features (and velocity-peaks) on the PV diagrams.
These all imply no significant contribution of the $x_1$-orbits within
the inner ILR to our PV diagram.
(2) Even if the gaseous $x_1$-orbits are elongated, they must be near
perpendicular to our line-of-sight, and cannot show us high velocities
which occur at their pericenter,
because the position angles $\psi$ of $x_1$- and $x_2$-orbits, measured
clockwise from the bar to the north (Figure \ref{fig:orbits} {\it top}),
are intrinsically $\psi_{x_1}=0\arcdeg$ and $\psi_{x_2}=90\arcdeg$
respectively for collisionless stars.
Owing to the damping force (working as viscosity), the two orbits drag
each other and make $\psi_{x_1}>0\arcdeg$ and  $\psi_{x_2}<90\arcdeg$
for collisional gas. This mechanism does not allow $\psi_{x_1}$ to become
larger than $\psi_{x_2}$. In addition to $\psi_{x_1}<\psi_{x_2}$,
our gaseous $x_2$-orbits (nuclear disk) of NGC 3079 have to be inclined
southward (counterclockwise) on the sky from our line-of-sight
for the reason discussed in $\S$ \ref{sec:orgndsk}.
Thus the elongated gaseous $x_1$-orbits (if they
exist) must be near perpendicular to our line-of-sight, and cannot show
the observed high-velocity features.

There are many disk galaxies which exhibit double bars, i.e. global and
nuclear bars \citep[e.g.][]{sw95, fr96, es99}. If NGC 3079 has a nuclear
bar other than our global bar, it may result in the observed high-velocity
features. This however is not the likely origin because such a nuclear
bar is also thought to be a consequence of the $x_1$- and $x_2$-orbits
in our global bar, whose orientations cannot be aligned with our
line-of-sight. Recently \citet{hs01} discussed that a nuclear bar could
be dynamically decoupled with a global bar in an evolutional phase,
having a random orientation from the global bar. In that phase
the $x_1$- and $x_2$-orbits have dissapeared \citep{hs01}, which is
not the case for NGC 3079.

Central components with high velocities are often found in rotation-curve
studies of external galaxies \citep{sf99}. They are attributed to central
massive components \citep{sf99, ts00} while gas streaming motions in a bar
could be another possible origin \citep{sk99}. In this section, we clarified
detailed kinematics and the effect of a bar on the gas disk of NGC 3079,
and discussed that the high velocity should be attributed to a massive
component at the galaxy center.

\section{CENTRAL ROTATION CURVE AND MASSIVE CORE}\label{sec:dms}

We discussed in \S \ref{sec:orgcore} that the high velocity peaks
at the radius of $\sim1\arcsec$ in the PV diagram are not likely
to result from noncircular motions driven by a bar, but are
a consequence of circular rotation around the nuclear core.
We here estimate the central dynamical mass using a rotation curve.

\subsection{Central Rotation Curve}

The observed PV diagram in the galaxy center suffers from large intrinsic
velocity dispersions, and spill-in of emission from outer low
velocity components owing to the finite spatial resolution and slit
width. Thus the central rotation curve is not straightforwardly obtained
from an observed PV diagram. We adopt the Takamiya \& Sofue method
\citep[TS method; Takamiya \& Sofue in private communication, and see]
[]{sr01} to obtain
the rotation curve of the nuclear core. This method is designed to
iteratively determine a rotation curve in an innermost region of
a galaxy so that a model PV diagram calculated from the rotation
curve and the galaxy's emission profile could reproduce the original
diagram.

Figure \ref{fig:tsrc} displays PV diagrams and rotation curves of
the nuclear core: the left-hand panel shows the observed one
along the galaxy major axis with $1\arcsec$ slit-width,
while the middle and right-hand panels show the calculated ones
in the TS method and in the peak-tracing method respectively.
The emission profile at the center ($\sim1\arcsec$) cannot be
derived directly from our data owing to our spatial resolution
($\sim2\arcsec$). But the observations are indicative of its very
sharp gradient, because even a point source can reproduce our
observed profile through a convolution with the synthesized beam.
Thus we assumed the emission profile as an analytic function of
an exponential with the scale length of $1\arcsec$, which
well reproduces observed features on the PV diagram.
We assumed intrinsic velocity dispersions of $60\kmps$
\citep[e.g. $15-50\kmps$ in the Galactic center from][which
is less active than NGC 3079]{ba88}.

\placefigure{fig:tsrc}

The PV diagram and rotation curve from the TS method show
that the envelope of the observed PV diagram requires a sharp
rise of the rotation curve at the central region. The observed
emission at around $0\kmps$, which is not well reproduced in
the TS method, may be contaminated by the emission from the
nuclear disk surrounding the nuclear core.
For a comparison, the right-hand panel displays the PV diagram
using the rotation curve derived by tracing peak-intensity
velocities of the observed PV diagram (the peak-tracing method);
this method is often used to obtain rotation curves, but fails
in reproducing the observed high velocity at the central part.

\subsection{Mass Estimation and Central Massive Core}

The dynamical mass within a radius $r$ is estimated by
\begin{equation}
M_{\rm dyn} =
2.3 \times 10^5 \left( \frac{r}{\rm kpc} \right)
\left( \frac{v(r)}{\rm km \, s^{-1}} \right)^2 \Msun.
\end{equation}
Our spatial resolution is about $2\arcsec$ ($152\pc$), and the
rotation velocity at this radius is $220\kmps$ from the rotation
curve in the TS method (Figure \ref{fig:tsrc} {\it middle}).
Thus the dynamical mass within the central $152\pc$ of NGC 3079 is
estimated to be $M_{\rm dyn}=1.7\times10^9\Msun$. The ratio of
dynamical mass to the total molecular gas mass ($3.0\times10^8\Msun$
from \S \ref{sec:core}) in the nuclear core is
$M_{\rm gas}/M_{\rm dyn} = 18\%$.

The rotation curve is meaningful at the radius less than $2\arcsec$,
even though the spatial resolution is only $\sim2\arcsec$:
We can distinguish two emission peaks only if their spatial separation
is larger than $2\arcsec$ on a single channel map or zeroth-moment map.
However on the PV diagram and rotation curve, the additional parameter,
velocity, provides a finer spatial resolution, because two emission
peaks with different velocities can be resolved in the PV diagram
even if their separation is less than $2\arcsec$.
Then the spatial resolution will be determined by the typical
error of a gaussian fit to a point source, which depends on
the signal-to-noise ratio, and in our case is $\ll 1\arcsec$
\citep[see][]{co97}. Therefore the velocity of $200\kmps$ at
the radius of $1\arcsec$ ($76\pc$) is meaningful, and implies
a dynamical mass of $M_{\rm dyn}=7.0 \times 10^8\Msun$ within
the radius of $76\pc$. The derived masses are listed in Table
\ref{tab:masstab}.

Recently central supermassive black holes with mass $10^{7-8}\Msun$
have been found in many AGNs \citep{mi95, wa99, is01}. In the case
of NGC 3079, VLBI observations of $\rm H_2O$ maser emission indicate
that its central mass within $0.5\pc$ is $\sim10^6\Msun$ \citep{tr98}.
Thus our mass $\sim10^9\Msun$ within $\sim100\pc$ cannot be attributed
to this central supermassive black hole. The connection between 
this large dynamical mass and the nuclear outflow in parsec-scale
\citep{hk84, is88, tr98} and kiloparsec-scale \citep{ds88,fd86} in
this galaxy will be an interesting issue for future investigations.

\section{CONCLUSIONS}\label{sec:conclusion}

We have made CO (1-0) observations of the $\Halpha$/Radio lobe
galaxy NGC 3079 with the Nobeyama Millimeter Array, and reported
characteristic features and gas dynamics in the molecular
disk in the central $4.5\kpc$ ($1\arcmin$).

1.
Our observations show four distinct components in the molecular
disk: the main disk, spiral arms, nuclear disk, and nuclear core.

2.
The main disk extends along the galaxy major axis. We detected the
inner region of this component, within the radius of $\sim 2\kpc$ on
a zeroth-moment map, while its full extent is not covered in our synthesis
observations. Molecular gas shows a smooth distribution on the main
disk, and has a gas mass of $5 \times 10^9\Msun$ within the central
$\sim 2\kpc$.

3.
The spiral arms are superimposed on the main disk, forming an
inverted $S$-shaped pattern on the sky.
Abrupt velocity changes of up to $\sim200\kmps$ are observed
along the spiral arms, in $S$-shaped twists of isovelocity contours
and double velocity-peaked features on the spectra.
This component and the nuclear disk form the so-called
``figure-of-eight'' pattern on a PV diagram, which sometimes appears
in edge-on barred galaxies.

4. 
The nuclear disk, appearing in PV diagrams, has an extent of
$\sim600\pc$ radius. The molecular gas is intensely concentrated
on the nuclear disk, where the gas surface density is an order of
magnitude higher than that typical of nearby galaxies.
Its appearance (bends and twist) on PV diagrams is indicative
of oval motions in the gas, rather than circular. And the
position angle of the oval orbits is inclined southward from
our line-of-sight.

5.
The nuclear core has a radius smaller than $\sim150\pc$, which is
unresolved in our current observations. The gas mass within the central
$\sim150\pc$ amounts to $3\times10^8\Msun$.
This component shows a very high velocity $\sim200\kmps$ at
the radius of the central $\sim100\pc$ on the PV diagram.

6.
A weak bar model explain the features of the main disk,
spiral arms, and nuclear disk in zeroth- and first-moment maps,
PV diagrams, and the spectrum at each spatial point.
The main disk and spiral arms result from gaseous $x_1$-orbits and
their crowding respectively. The nuclear disk arises from gaseous
$x_2$-orbits. The gas concentration in the nuclear disk is also
explained in the context of our model: the gas on $x_1$-orbits moves
along spiral arms (or offset shocks), collides with the gas on
$x_2$-orbits, and is accumulated onto the nuclear disk.
Assuming that the gas in spiral arms moves along streamlines
perpendicular to our line-of-sight, the pattern speed of the bar can be
estimated to be $55\pm10 \kmpspkpc$.

7.
Our model for a bar however does not explain the high velocity of
the nuclear core. Moreover, any orbit caused by a bar is not
likely to produce that component. Thus we attribute it to a central
massive core with a dynamical mass of $10^9\Msun$
within the central $100\pc$, which is three orders of magnitude
more massive than the mass of a central supermassive black hole.

\acknowledgments
We are grateful to the NRO staff and Rainbow team members for their
help with observations. We thank Mr. Toshihito Shibatsuka for
kindly showing his results prior to publication. We also thank
the referee for useful comments. J. K. thanks Dr. Toshihiro Handa
for careful reading of the manuscript and fruitful comments,
Dr. Keiichi Wada for useful discussions and continuous encouragements,
and Dr. Kentaro Aoki for his help with the analysis of
the {\it HST} data. J. K. was financially supported by the Japan
Society for the Promotion of Science for Young Scientists.

\begin{deluxetable}{lcc}
\tablecolumns{3} 
\tablewidth{0pc} 
\tablecaption{PROPERTIES OF NGC 3079}
\tablehead{
\colhead{Parameter} & \colhead{Value} & \colhead{Reference}}
\startdata
Morphology ........................                & SB(s)c sp         & 1 \\
                                                    & Sc pec:           & 2 \\
Nuclear activity .................                & LINER             & 3 \\
                                                    & Type 2 Seyfert    & 4 \\
Position of nucleus:                                &                   & 5 \\
$\,\,\,\,\,\,\,\,\,\,\alpha$(B1950) ....................& $9\,\,58\,\,35.02$&   \\
$\,\,\,\,\,\,\,\,\,\,\delta$(B1950) .....................& $55\,\,55\,\,15.4$&   \\

$D_{25} \times d_{25}$ ..........................& $7.9 \times 1.4$  & 1 \\
$B_{T}$(mag) ...........................          & 11.54             & 1 \\
P.A. ($\arcdeg$) ............................                  & 165               & 1 \\
Inclination ($\arcdeg$) ...................                  &  77               & 6 \\
Distance (Mpc) .................                       & 15.6              & 5 \\
Linear scale ($\rm pc\,\,arcsec^{-1}$)          & 76                &   \\
$S_{60 \rm \mu m}$ (Jy) ........................& 50.17 $\pm 0.054$ & 7 \\
$S_{100 \rm \mu m}$ (Jy) .......................& 103.40 $\pm 0.154$& 7 \\
$L_{\rm FIR}$ ($\Lsun$)\tablenotemark{a}..............................& $2.1\times10^{10}$&   \\
\enddata
\tablenotetext{a}{$L_{\rm FIR}$ was calculated as
$3.75 \times 10^5 (D/\mpc)^2 (2.58 S_{60 \rm \mu m} + S_{100 \rm \mu m})$
in units of $\Lsun$.}
\tablecomments{
Reference.--
(1) de Vaucouleurs et al. 1991;
(2) Sandage \& Tamman 1981;
(3) Heckman 1980;
(4) Ford et al. 1986;
(5) Sofue \& Irwin 1992;
(6) Young et al. 1995;
(7) Soifer et al. 1989}
\label{tab:prop}
\end{deluxetable}

\begin{deluxetable}{lc}
\tablecolumns{2} 
\tablewidth{0pc} 
\tablecaption{PARAMETERS OF CO (1-0) CUBE}
\tablehead{
\colhead{Parameter} & \colhead{Value}}
\startdata
Configuration .......................            & AB+C+D \\
Weighting ............................           & NA     \\
Field of view ($\arcsec$) ..................     & 65     \\
Synthesized beam:                                &        \\
$\,\,\,\,\,\,\,\,\,\,\,\,$major axis ($\arcsec$) ..............  & 1.9    \\
$\,\,\,\,\,\,\,\,\,\,\,\,$minor axis ($\arcsec$) ..............  & 1.6    \\
$\,\,\,\,\,\,\,\,\,\,\,\,$P.A. ($\arcdeg$) ....................  & 105    \\
Velocity resolution ($\kmps$)                    & 10.4   \\
rms ($\rm mJy\,beam^{-1}$)................       & 12     \\
\enddata
\label{tab:cube}
\end{deluxetable}

\begin{deluxetable}{lc}
\tablecolumns{2} 
\tablewidth{0pc} 
\tablecaption{KINEMATIC PARAMETERS OF THE MAIN DISK}
\tablehead{
\colhead{Parameter} & \colhead{Value}}
\startdata
Dynamical center:                                   &                       \\
$\,\,\,\,\,\,\,\,\,\,\,\,\alpha$(B1950) .................& $9\,\,58\,\,35.00$    \\
$\,\,\,\,\,\,\,\,\,\,\,\,\delta$(B1950) .................& $55\,\,55\,\,15.90$   \\
P.A (deg) .......................                       & 169                   \\
Inclination (deg) ............               &  79                   \\
$\vsys$ ($\kmps$) .................            & 1147                  \\
\enddata
\label{tab:kin}
\end{deluxetable}

\begin{deluxetable}{ccccc}
\tablecolumns{5} 
\tablewidth{0pc} 
\tablecaption{GAS AND DYNAMICAL MASSES}
\tablehead{
\multicolumn{2}{c}{RADIUS\tablenotemark{a}} &  & \\
\multicolumn{2}{c}{------------------------} & \colhead{$M_{\rm dyn}$} & \colhead{$M_{\rm gas}$\tablenotemark{b}} & \colhead{$M_{\rm gas}/M_{\rm dyn}$}\\
\colhead{(arcsec)} & \colhead{(pc)} & \colhead{($10^{9}\Msun$)} & \colhead{($10^{8}\Msun$)} & \colhead{(\%)}}
\startdata
26 & $2.0\times10^3$ & $22$  & $46$   ($26$)  &  21 (12)\\
2  & $1.5\times10^2$ & 1.7   & $3.0$  ($1.7$) &  18 (10)\\
1  & 76   & 0.7 &   --       & --      \\
\enddata
\tablenotetext{a}{Radius in the galactic plane}
\tablenotetext{b}{Gas mass is calculated with a conversion factor $X_{\rm CO}=1.8\times10^{20}\unitcnv$ and
the H abundance $X=0.707$. The values in brackets are the ones with
$X_{\rm CO}=1.0\times10^{20}\unitcnv$ for high metallicity regions
such as galactic centers \citep{ar96}}
\label{tab:masstab}

\end{deluxetable} 

\clearpage

\begin{figure}
\epsscale{1.0}
\plotone{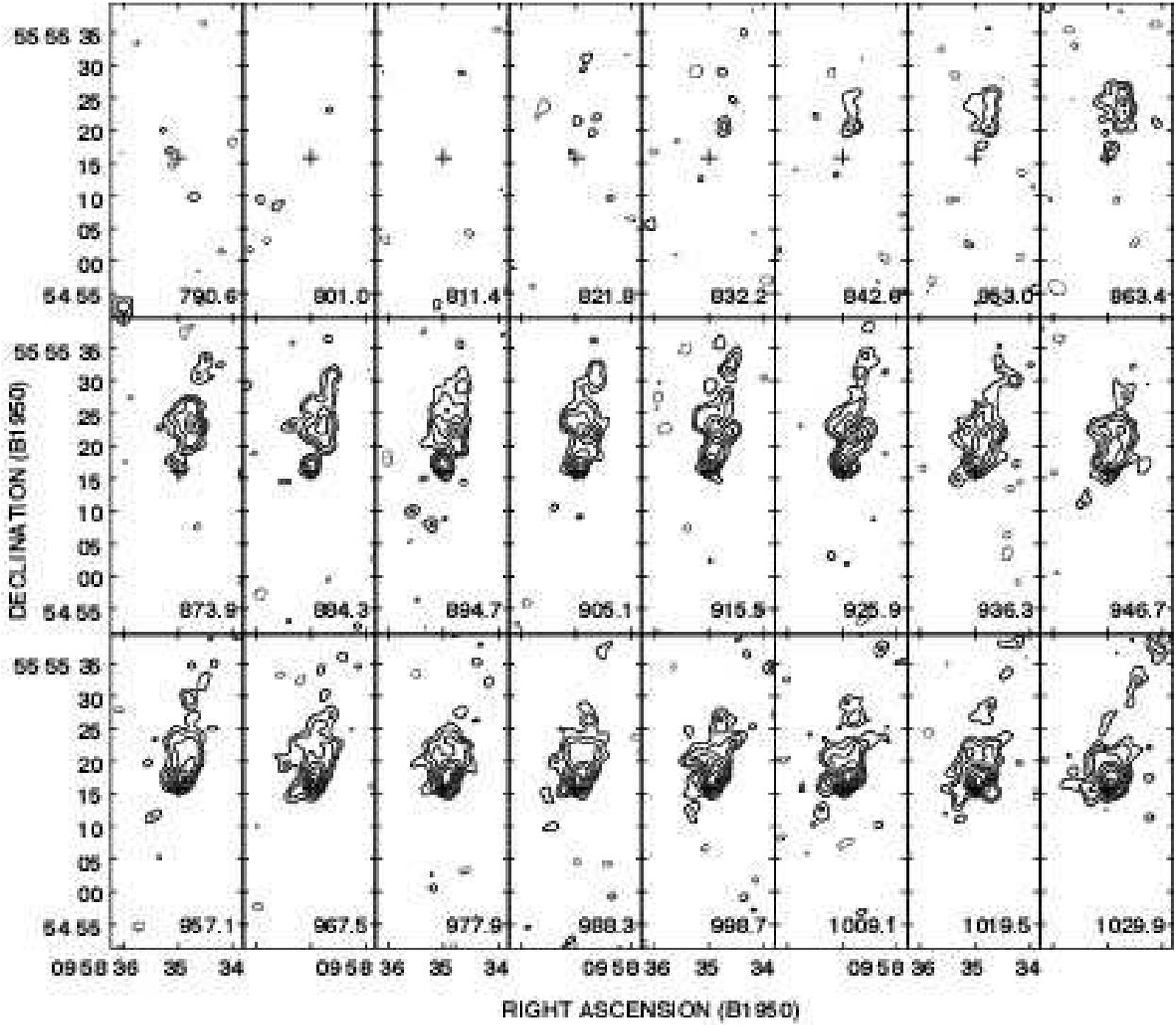}
\caption{Channel maps in CO (1-0) emission. The channels have an interval of
$10.4\kmps$. Their central velocities ($V_{\rm LSR}$ in $\kmps$) are
labeled at the lower-right corners. Crosses show the position of the dynamical
center (Table \ref{tab:kin}). Contour levels are -3, 3, 5, 8, 12, 16, 20,
25, 30$\sigma$, where $1 \sigma = 12 \mjypbm$. Negative contours are dotted.
No primary beam correction has been applied.
\label{fig:chmap}}
\end{figure}

\addtocounter{figure}{-1}
\begin{figure}
\epsscale{1.0}
\plotone{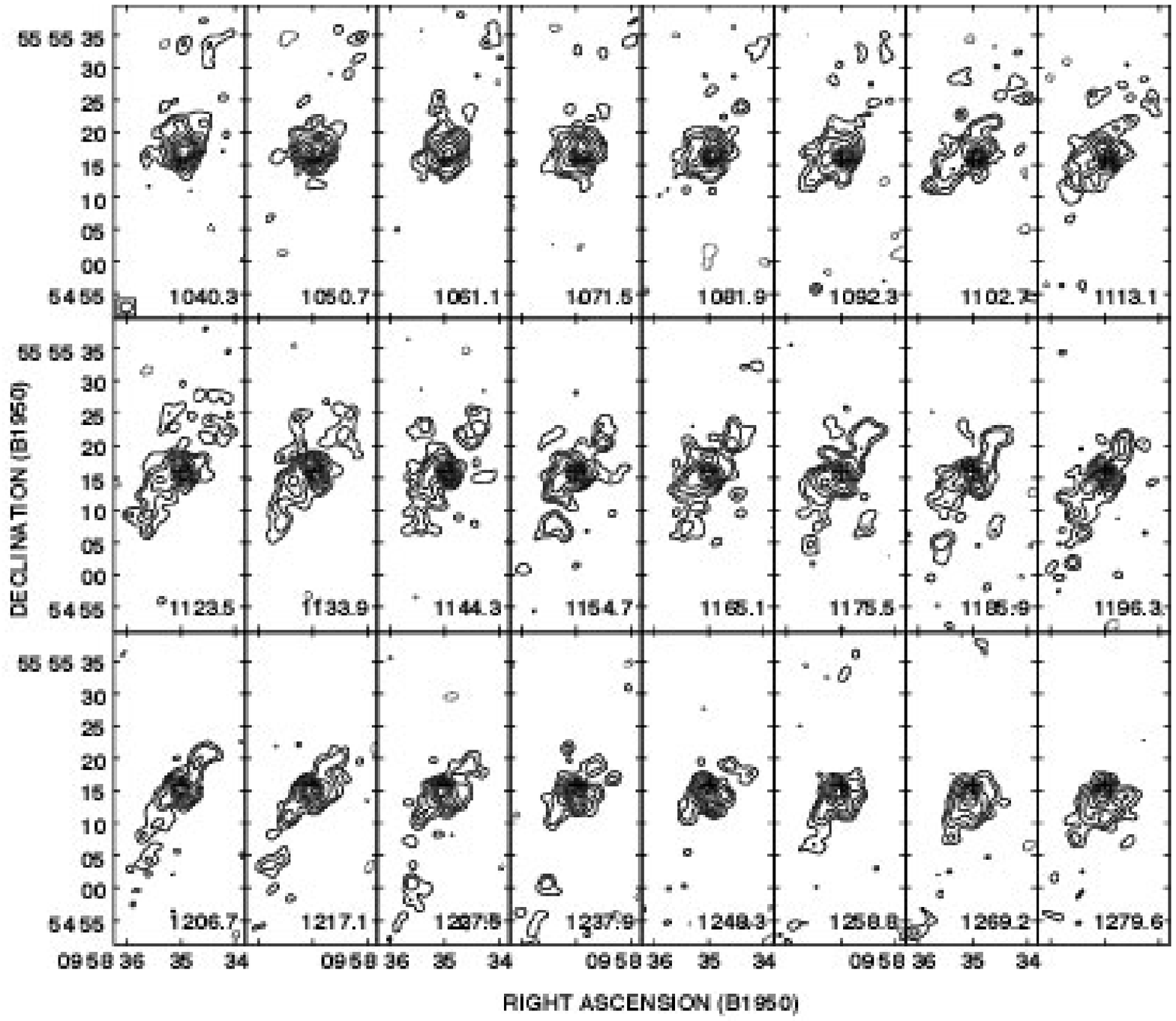}
\caption{continued.}
\end{figure}

\addtocounter{figure}{-1}
\begin{figure}
\epsscale{1.0}
\plotone{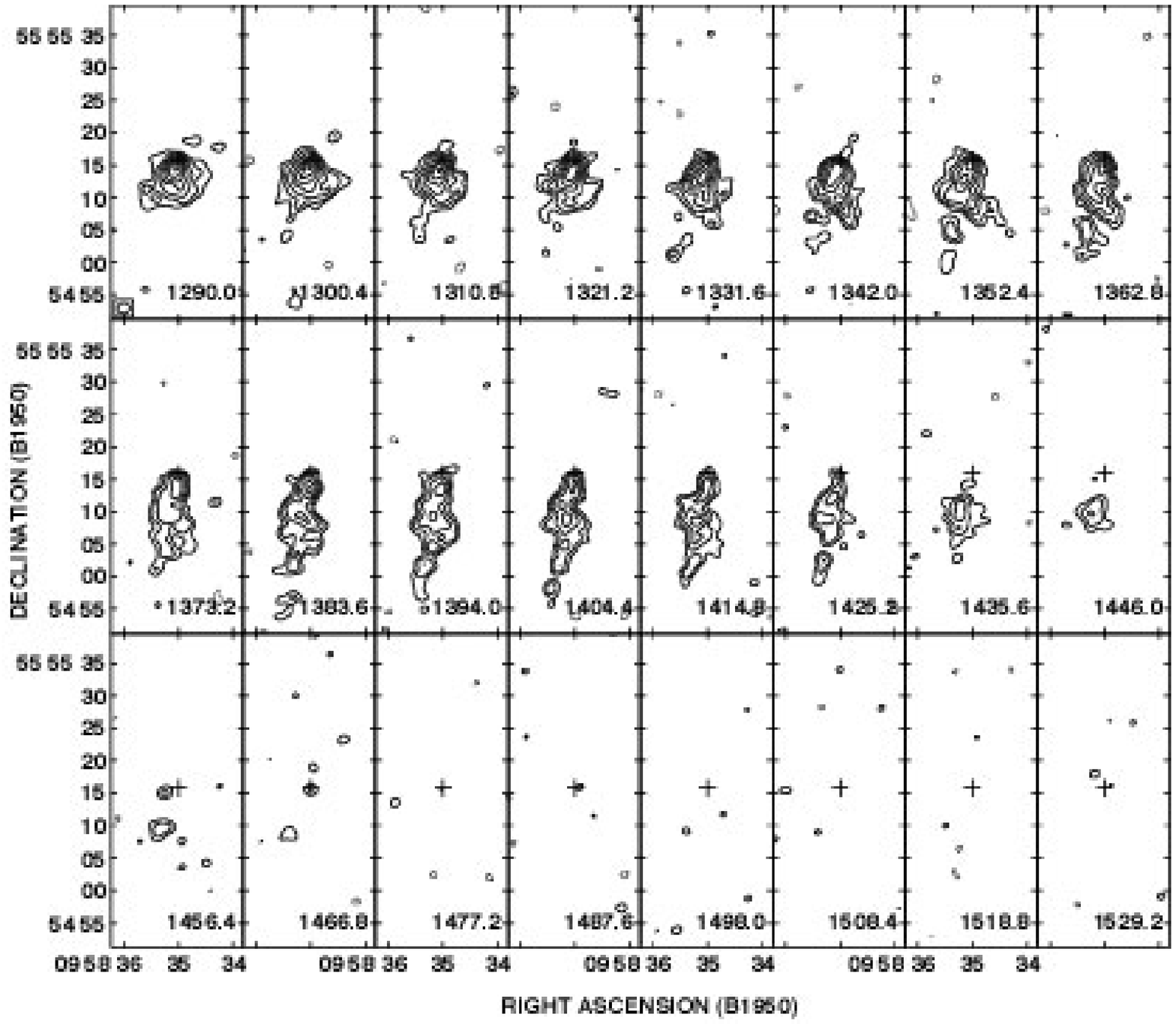}
\caption{continued.}
\end{figure}

\begin{figure}
\epsscale{1.0}
\plotone{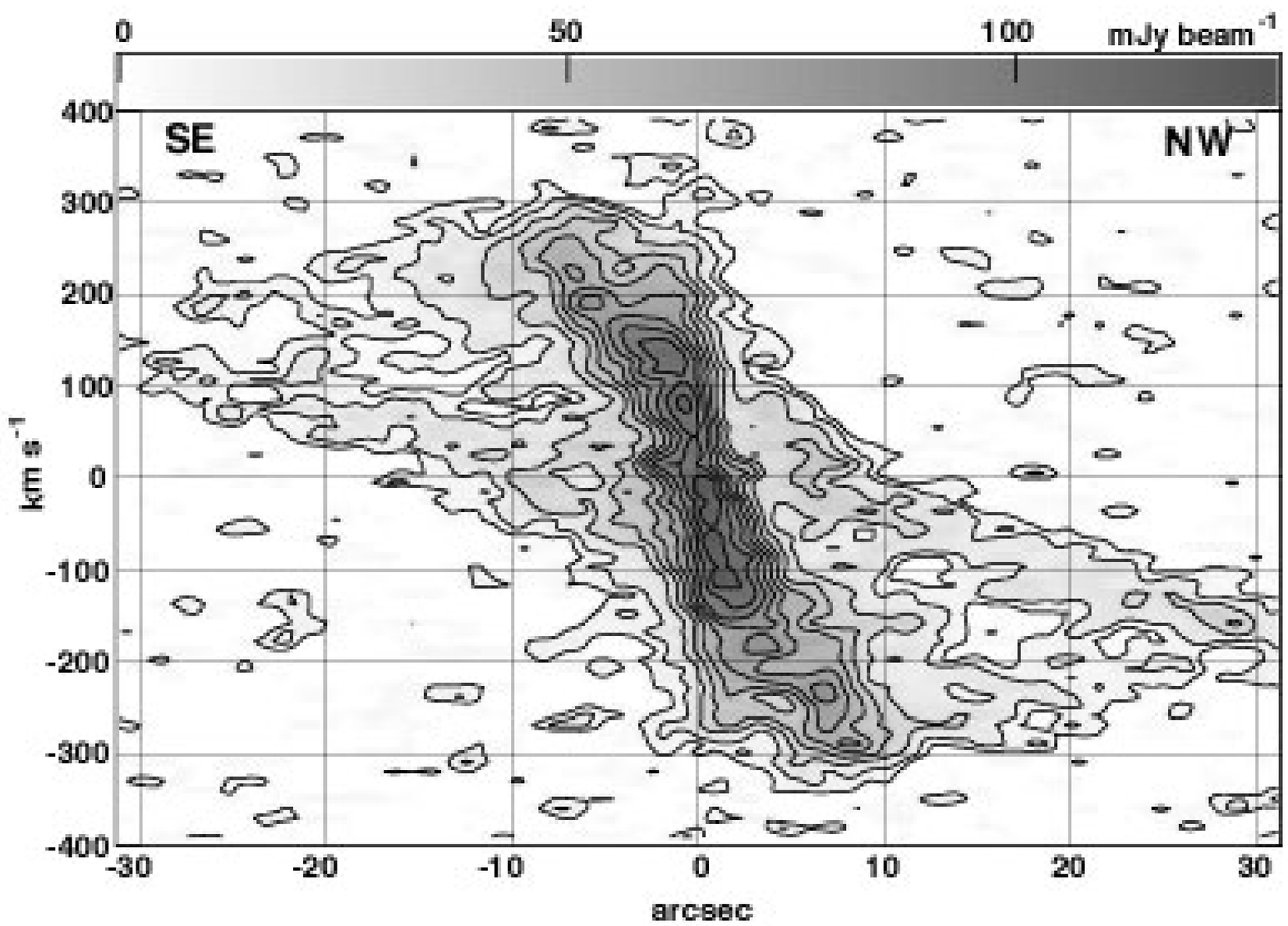}
\caption{PV diagram of CO (1-0) emission along the major axis of NGC 3079 (P.A.
$= 165\arcdeg$). The slit-width was set to be $12\arcsec$, containing almost
the entire emission in Figure \ref{fig:mom01}.
The axes are labeled relative to the dynamical center and
systemic recession velocity (Table \ref{tab:kin}). Velocities have been
corrected for inclination ($77\arcdeg$). Contours are at 5, 10, 20, 30, 40,
50, 60, 70, 80, 90, 100\% of the peak intensity.\label{fig:pvd}}
\end{figure}

\begin{figure}
\epsscale{0.6}
\plotone{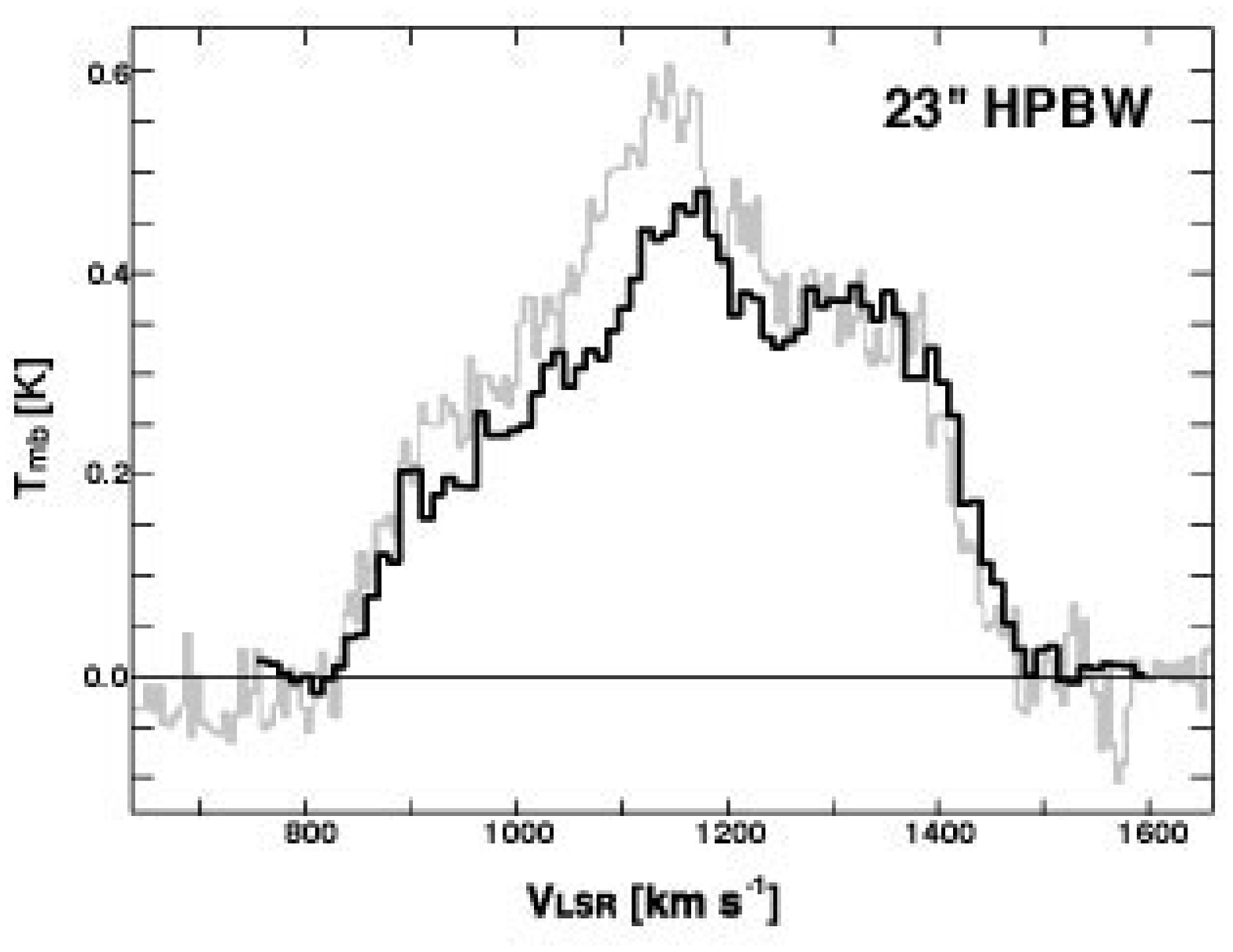}
\caption{Comparison of our CO (1-0) line spectrum with the single-dish result
of IRAM 30 m from \citet{br93}. The NMA cube was corrected for primary beam
attenuation, convolved with the single-dish beam size ($23\arcsec$) of IRAM,
and sampled at the pointing centers of the IRAM observations ($\alpha_{1950}$,
$\delta_{1950}$) = ($09^{\rm h}58^{\rm m}35^{\rm s}.4$, $+55\arcdeg55\arcmin11
\arcsec.0$).
The gray line show the spectrum from the IRAM observations, whereas the black
line show that from the NMA cube.\label{fig:spec}}
\end{figure}

\begin{figure}
\epsscale{1.0}
\plotone{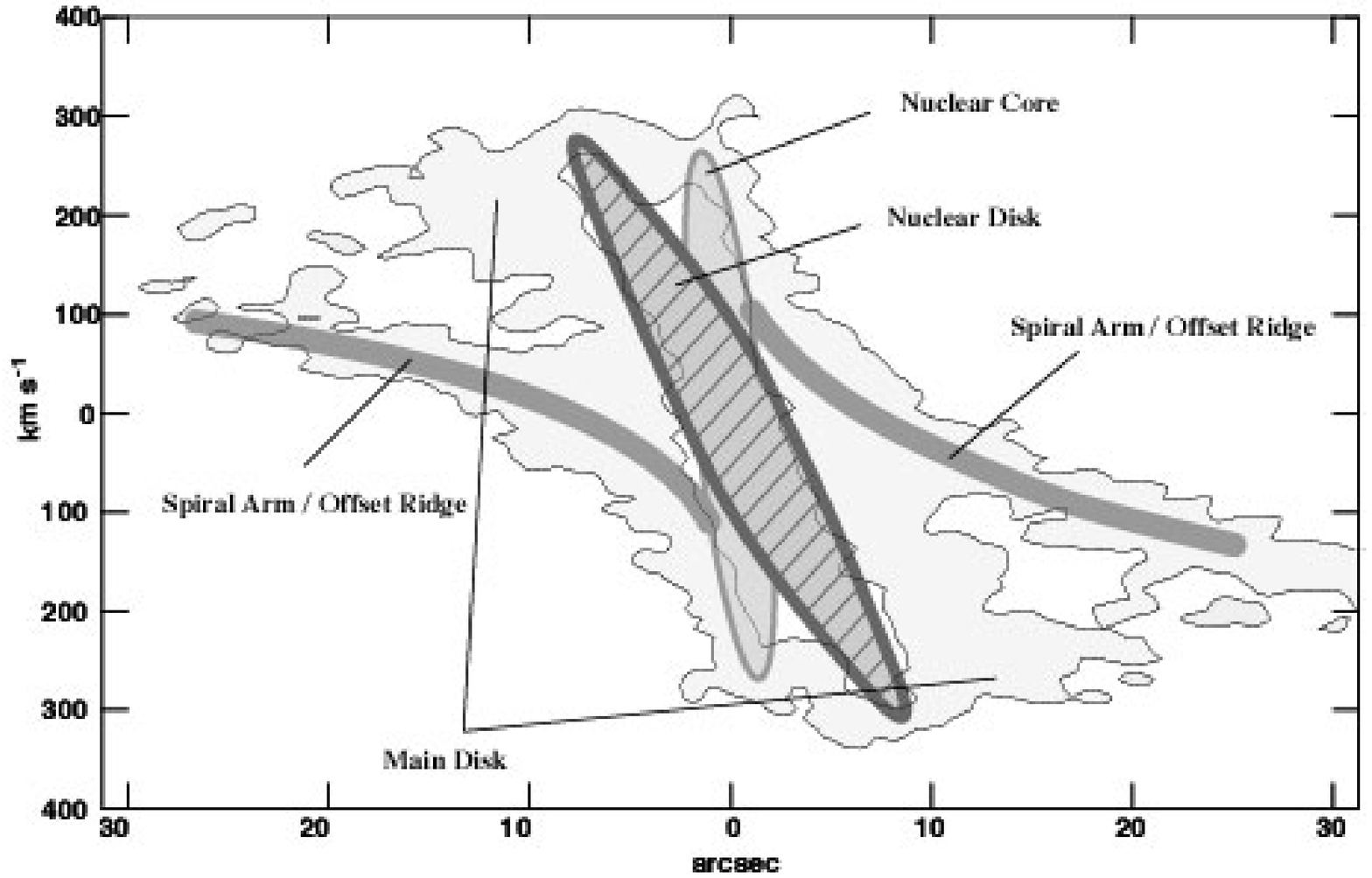}
\caption{Schematic illustration of PV diagram. Four distinct components
exist in the molecular disk of NGC 3079: main disk, spiral arms (or offset ridges),
nuclear disk, and nuclear core. \label{fig:schpvd}}
\end{figure}

\begin{figure}
\epsscale{1.0}
\plotone{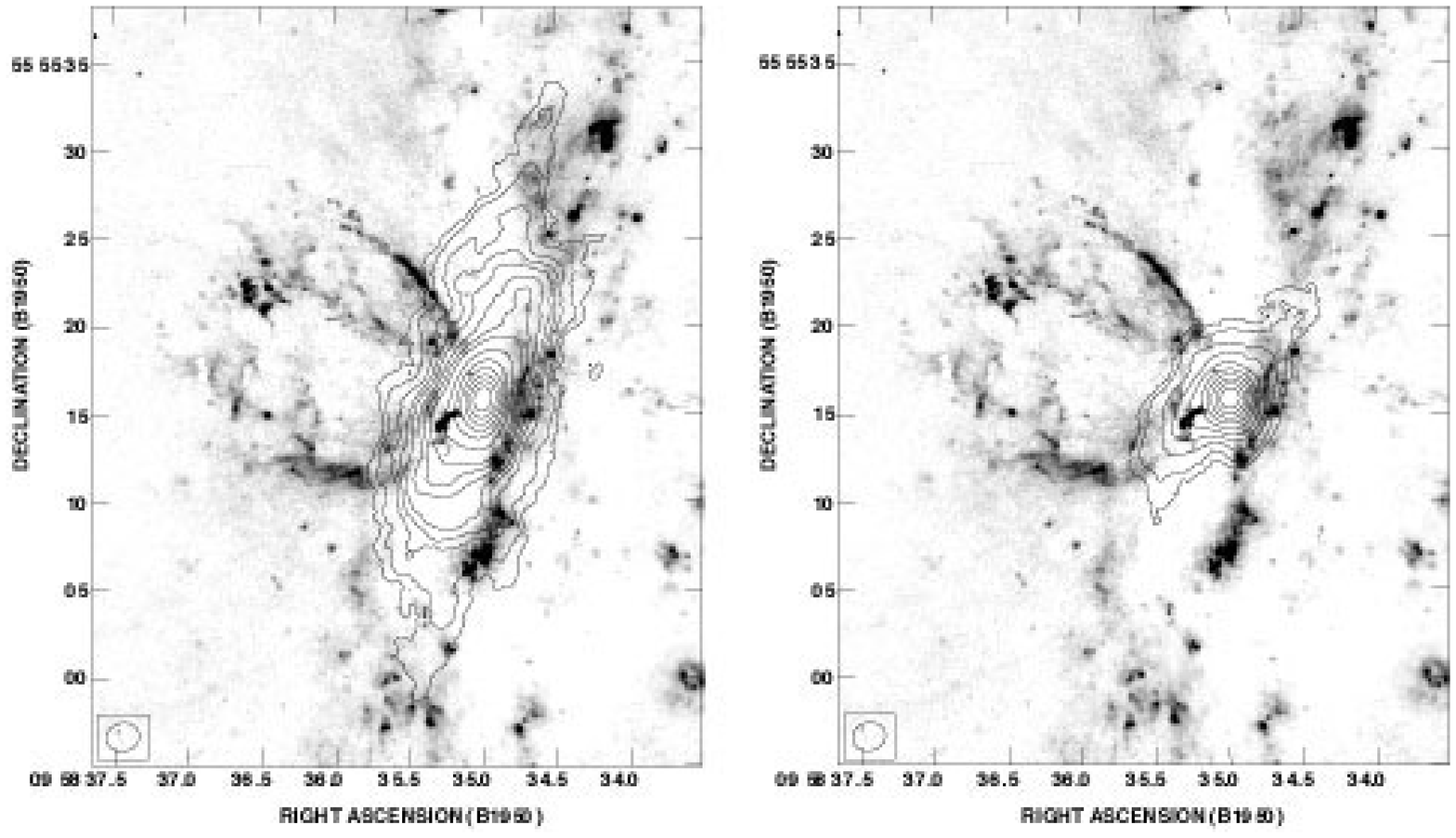}
\caption{Comparisons of CO maps (contours) with the $\Halpha+\NII$ map
(grayscale;
from the HST archives). The absolute position of the HST image was calibrated
with the USNO-A2.0 catalog, and is accurate to about $0.5\arcsec$. The ellipses
at the lower-left corners show the synthesized beam for the CO observations.
The two panels show two sets of CO zeroth-moment maps with different
integrated velocity ranges.
{\it Left:} The CO map (contours) is derived by integrating the cube in the
full velocity range where CO emission is detected. Contour levels are
3, 5, 9, 12, 15, 20, 25, 30, 40, 50, 60, 70, 80, 90, 100 \% of the peak
integrated intensity $1.25 \times 10^2 \jypbm \, \kmps$.
The molecular disk coincides with the void of HII regions at the center,
and its center is coincident with the root of the $\Halpha$ lobe.
{\it Right:} The CO map (contours) is derived by integrating the cube in the
velocity range where the spiral arms are dominant in emission
($\Delta V = \pm 100\kmps$). Contour levels are 5, 10, 20, 30, 40, 50,
60, 70, 80, 90, 100 \% of the peak integrated intensity $6.52 \times
10 \jypbm \, \kmps$. The spiral arms extend from the root of
the $\Halpha$ lobe, and wind counterclockwise.  \label{fig:haco}}
\end{figure}

\begin{figure}
\plotone{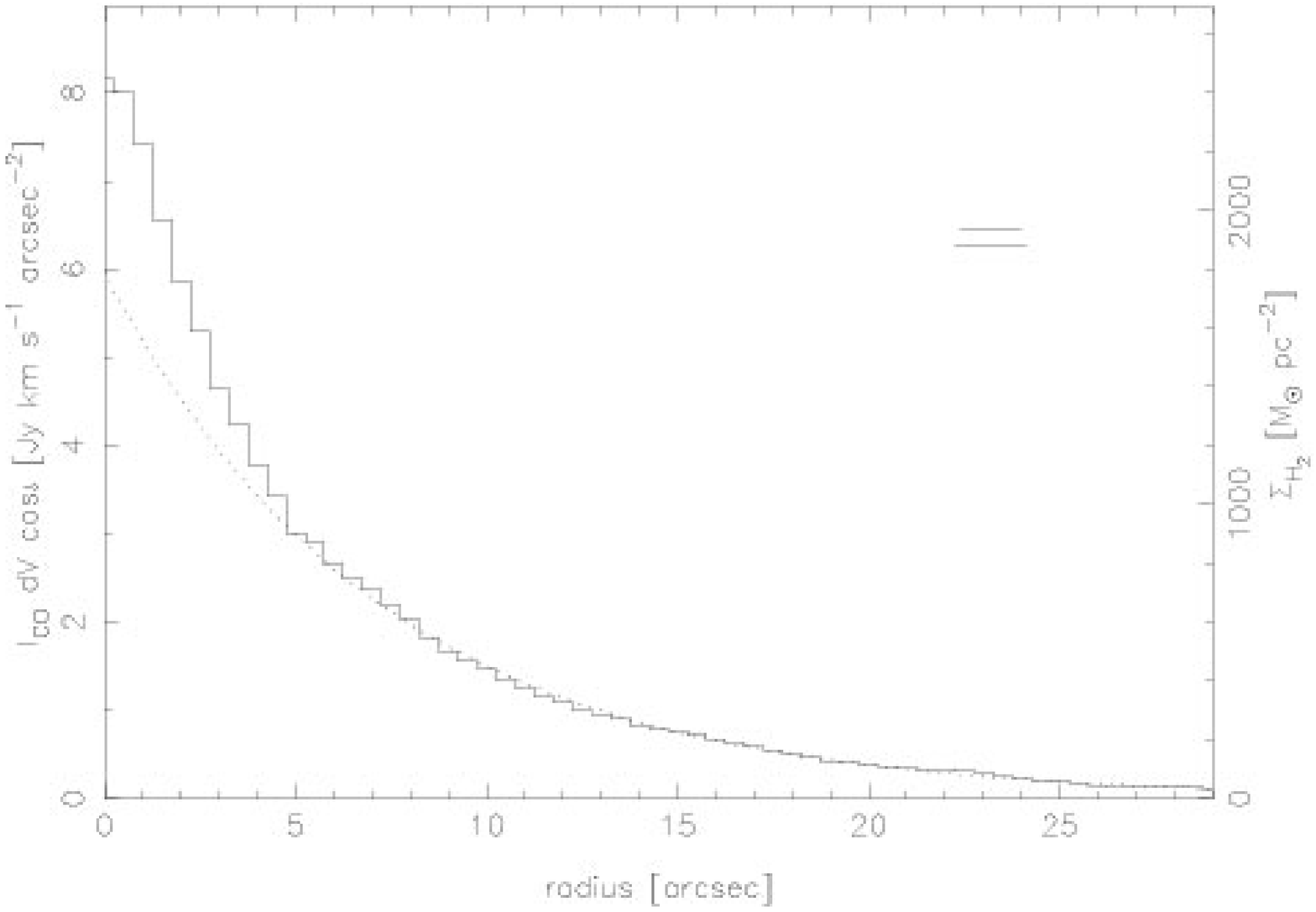}
\caption{Radial distribution of CO emission. CO integrated intensities are
azimuthally averaged with corrections for inclination and primary beam
attenuation. No correction for missing flux (about $10-30\%$) has been
applied. The surface densities on the right-hand axis are calculated
by assuming the Galactic conversion factor $X_{\rm CO} = 1.8 \times 10^{20}
\unitcnv$ \citep{dm01}.
The two horizontal bars show the beam size along the major and minor axes
($1\arcsec.9\times1\arcsec.6$). The solid line presents the derived emission
profile, while the dotted line shows a fitted exponential profile, i.e.
$5.97 \exp(-r/7\arcsec.24)$, at radii of $r>5\arcsec$.\label{fig:srfd}}
\end{figure}

\begin{figure}
\epsscale{1.0}
\plotone{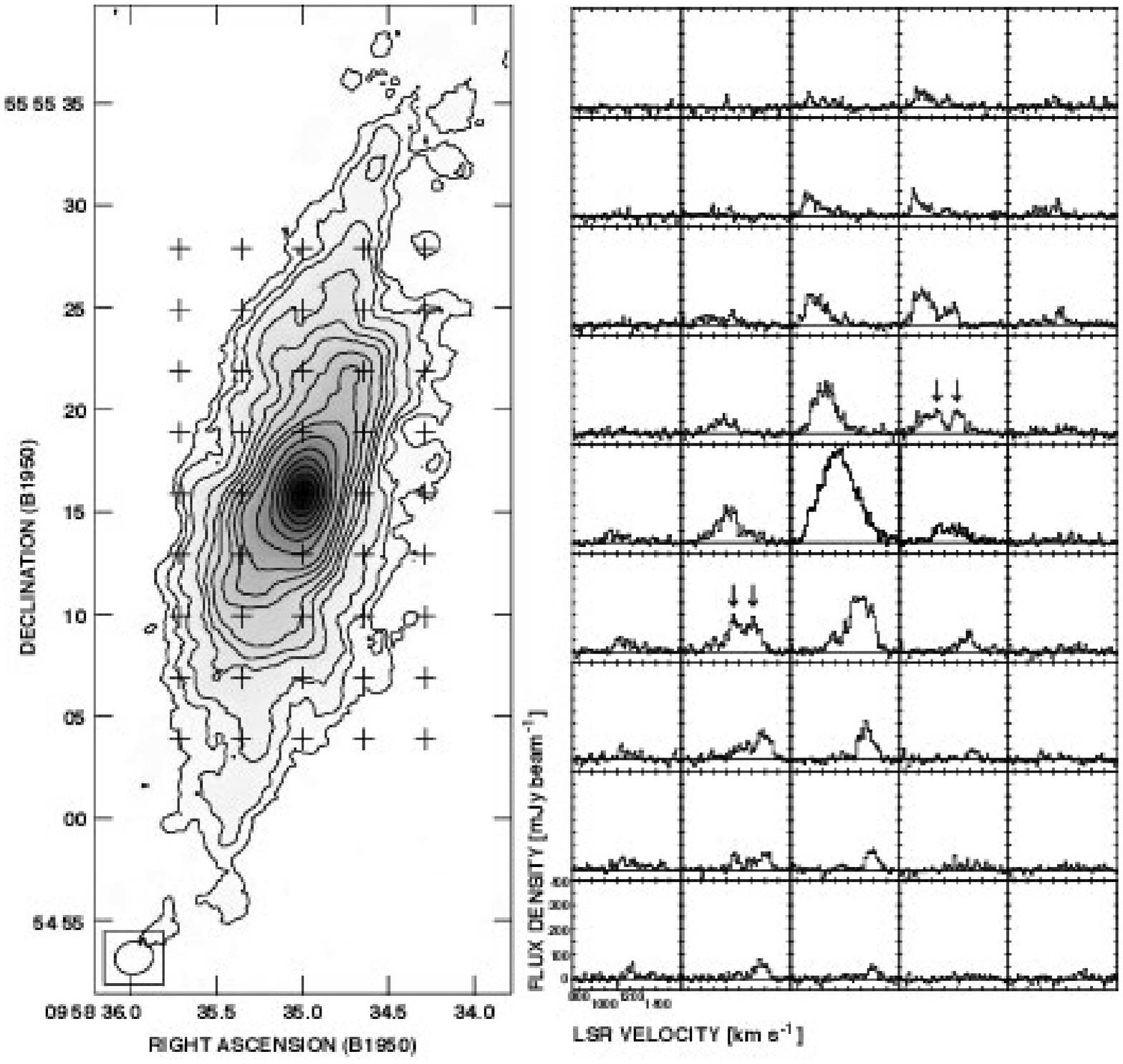}
\caption{CO (1-0) line profiles and zeroth-moment map for reference.
The line profiles are sampled at $5 \times 7$ points on the $3''$-spacing
grid (crosses on the left-hand panel), centered at the dynamical center
(Table \ref{tab:kin}). Double-peaked features in the line profiles are observed
along the spiral arms: arrows point to two examples. The double-peaks
originate from distinct velocities between the main disk and spiral arms
(or offset ridges).\label{fig:lineprof}}
\end{figure}

\begin{figure}
\epsscale{1.0}
\plotone{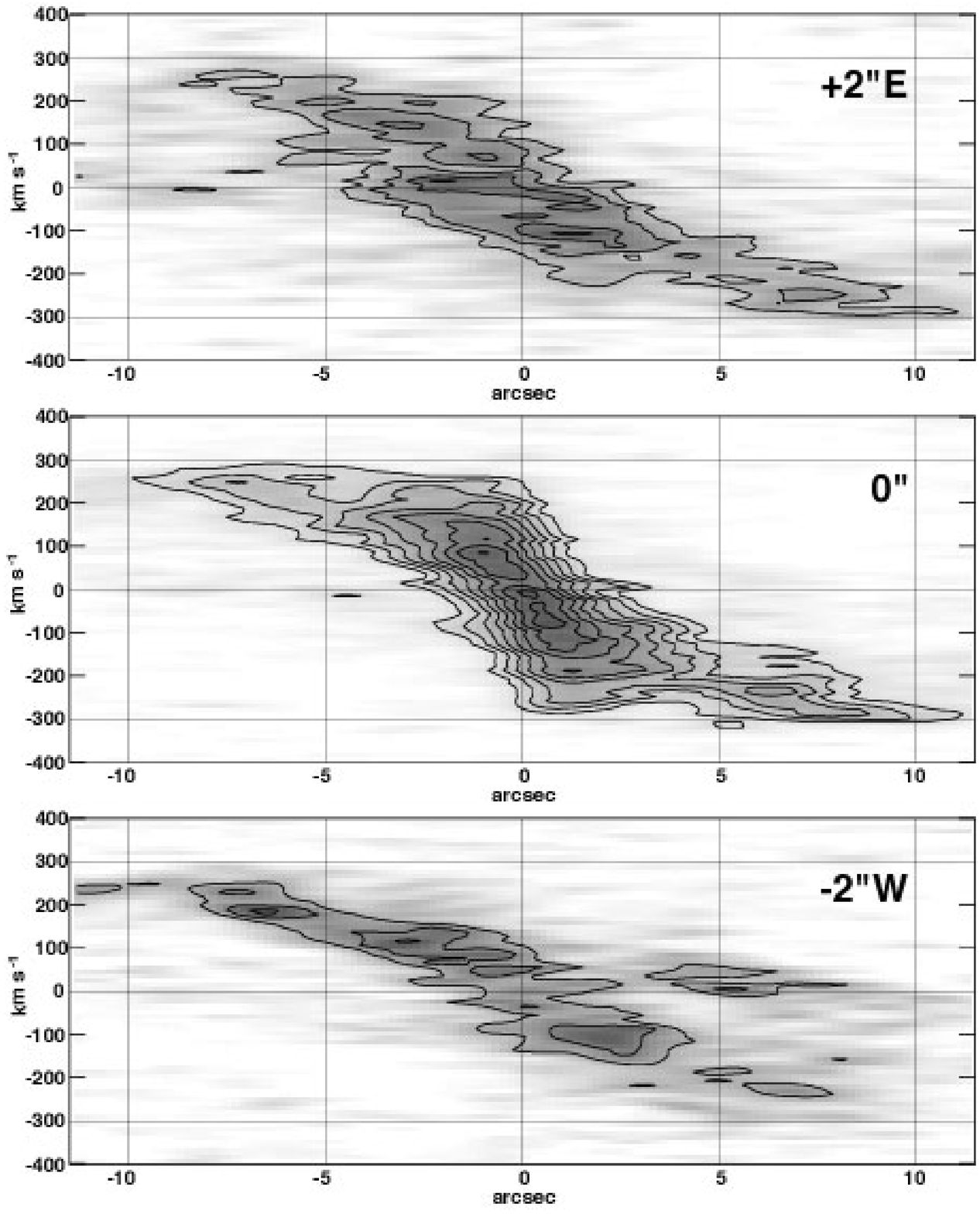}
\caption{
PV diagrams along the major axis (P.A. $= 165\arcdeg$) with offsets along the
minor axis of $+2\arcsec$, $0\arcsec$ and $-2\arcsec$ (``+'' denotes east).
The slit-widths are $1\arcsec$. The axes are labeled relative to the dynamical
center and systemic recession velocity (Table \ref{tab:kin}). Velocities have
been corrected for inclination ($77\arcdeg$). Contours in all the panels are
at 2, 3, 4, 5, 6, 7, 8, 9, 10 times $4.0 \times 10^{-2}\jypbm$.\label{fig:ewpvd}}
\end{figure}

\begin{figure}
\epsscale{0.8}
\plotone{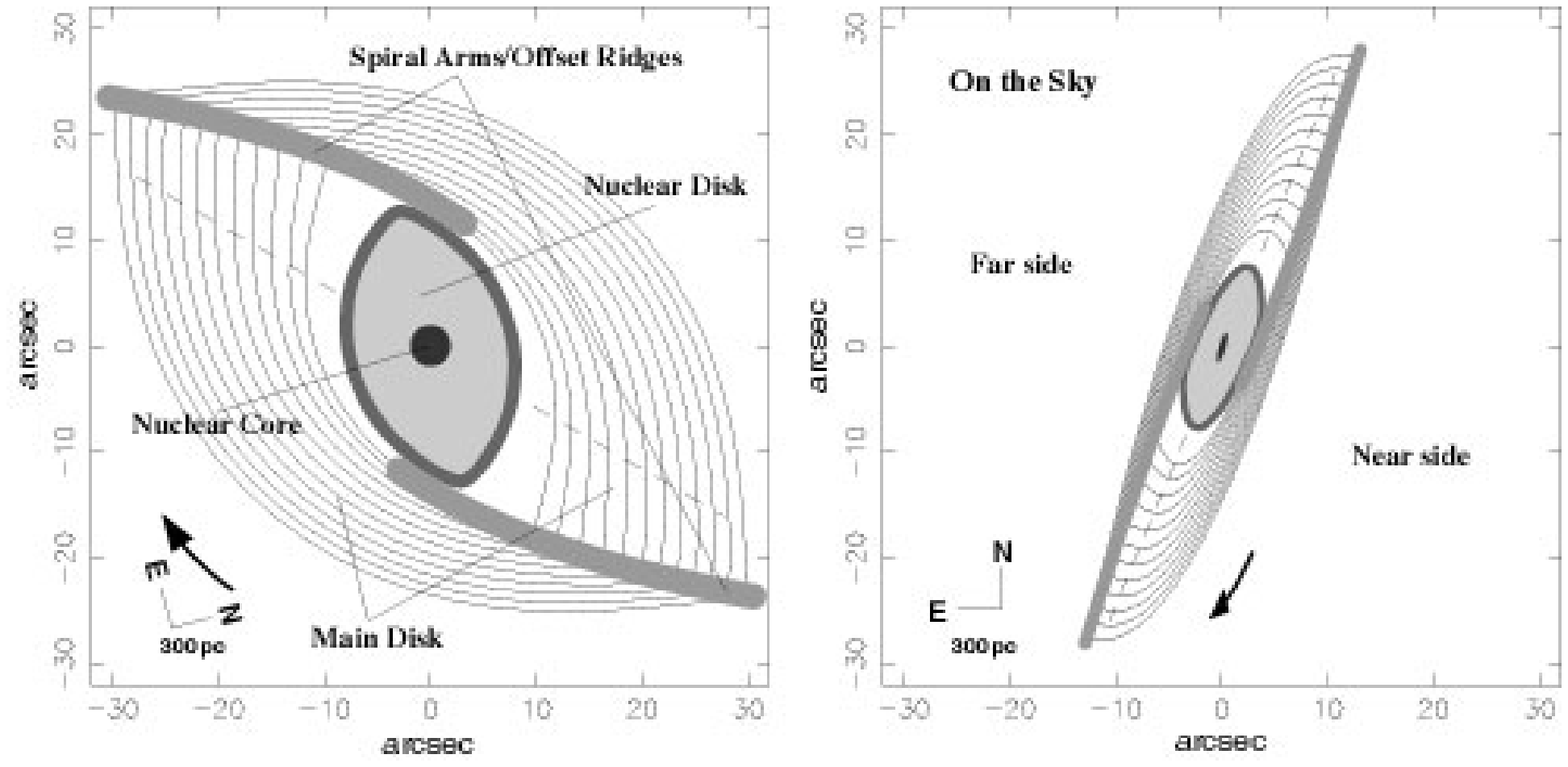}
\caption{
Schematic views of our interpretation of the molecular disk in NGC 3079.
Four distinct components, the main disk, spiral arms/offset ridges,
nuclear disk, and nuclear core, are drawn on the streamlines of our model
orbits in the bar reference frame (\S \ref{sec:dom}). The bar has the position
angle of $135\arcdeg$ intrinsic to the galaxy from the north.
{\it Left:} Face-on view. The galaxy major axis lies horizontally, while
the bar runs along dashed line which is inclined clockwise by $30\arcdeg$
from the horizontal.
{\it Right:} Projection on the sky for ${\rm P.A.}=165\arcdeg$ and
$i=77\arcdeg$.\label{fig:schview}}
\end{figure}

\begin{figure}
\epsscale{0.4}
\plotone{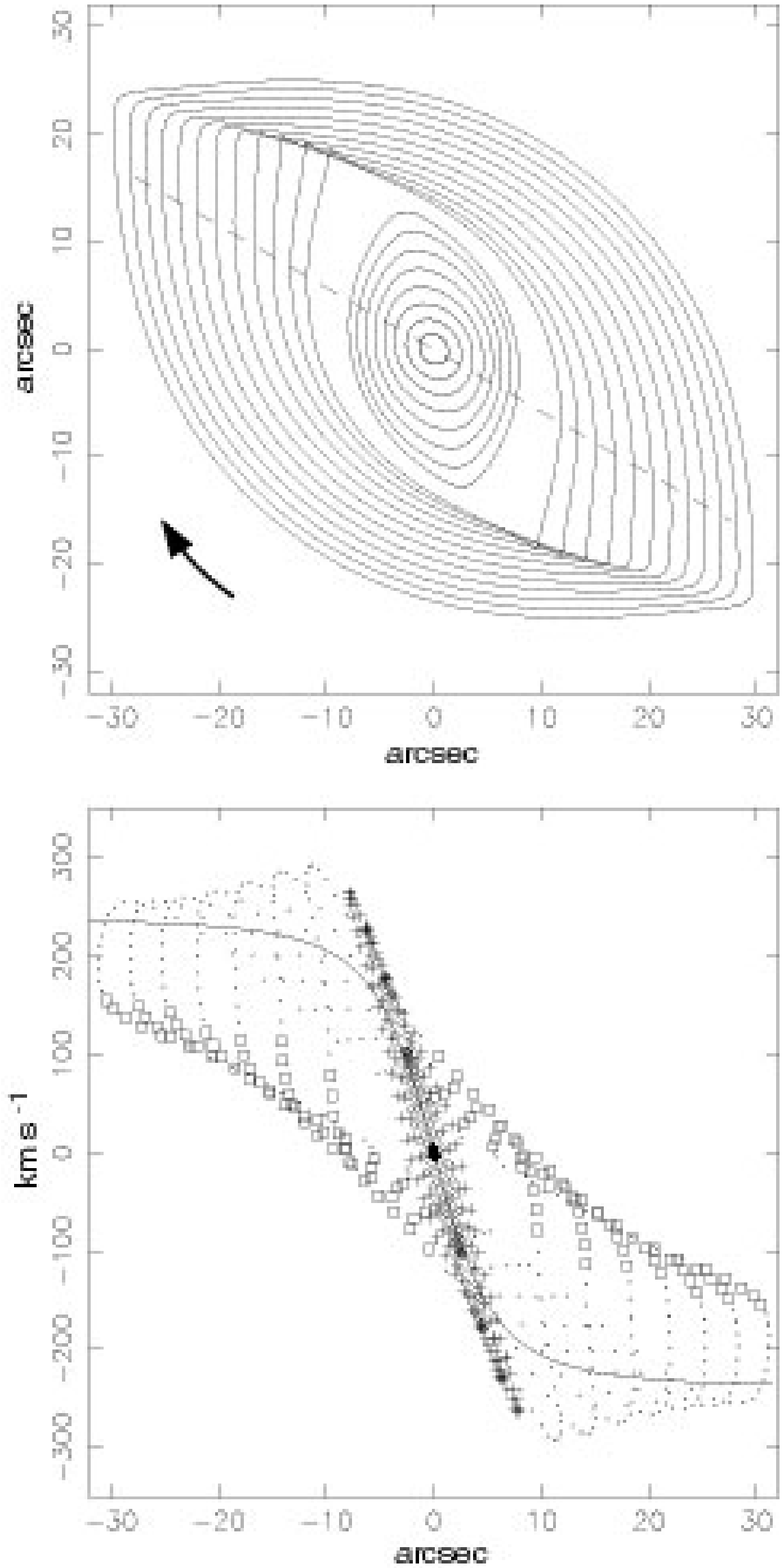}
\caption{Model gas motions in the molecular disk of NGC 3079.
{\it Top:} Face-on view of model orbits in a weak bar potential, on the
frame rotating with the bar. The galaxy major axis runs horizontally.
The bar is rotating clockwise, and runs along the dashed line (inclined
from the horizontal by $30\arcdeg$).
The orbits are calculated using a damped orbit model \citep{wa94}.
The inner and outer ILR and corotation occur at the radii of $8\arcsec$,
$11\arcsec$ and $57\arcsec$ ($1\arcsec=76\pc$) respectively.
{\it Bottom:} Model PV diagram, which cuts the model galaxy along the
galaxy major axis, with the width containing the full region at the top
panel. The solid line shows the circular rotation curve. Different symbols
are used for different orbits: circles are $x_1$-orbits inside the inner
ILR, crosses are $x_2$-orbits, squares are $x_1$-orbits outside the outer
ILR in front of the sharp turns on the streamlines, and dots are also
the $x_1$-orbits at the rests.
\label{fig:orbits}}
\end{figure}

\begin{figure}
\epsscale{1.0}
\plotone{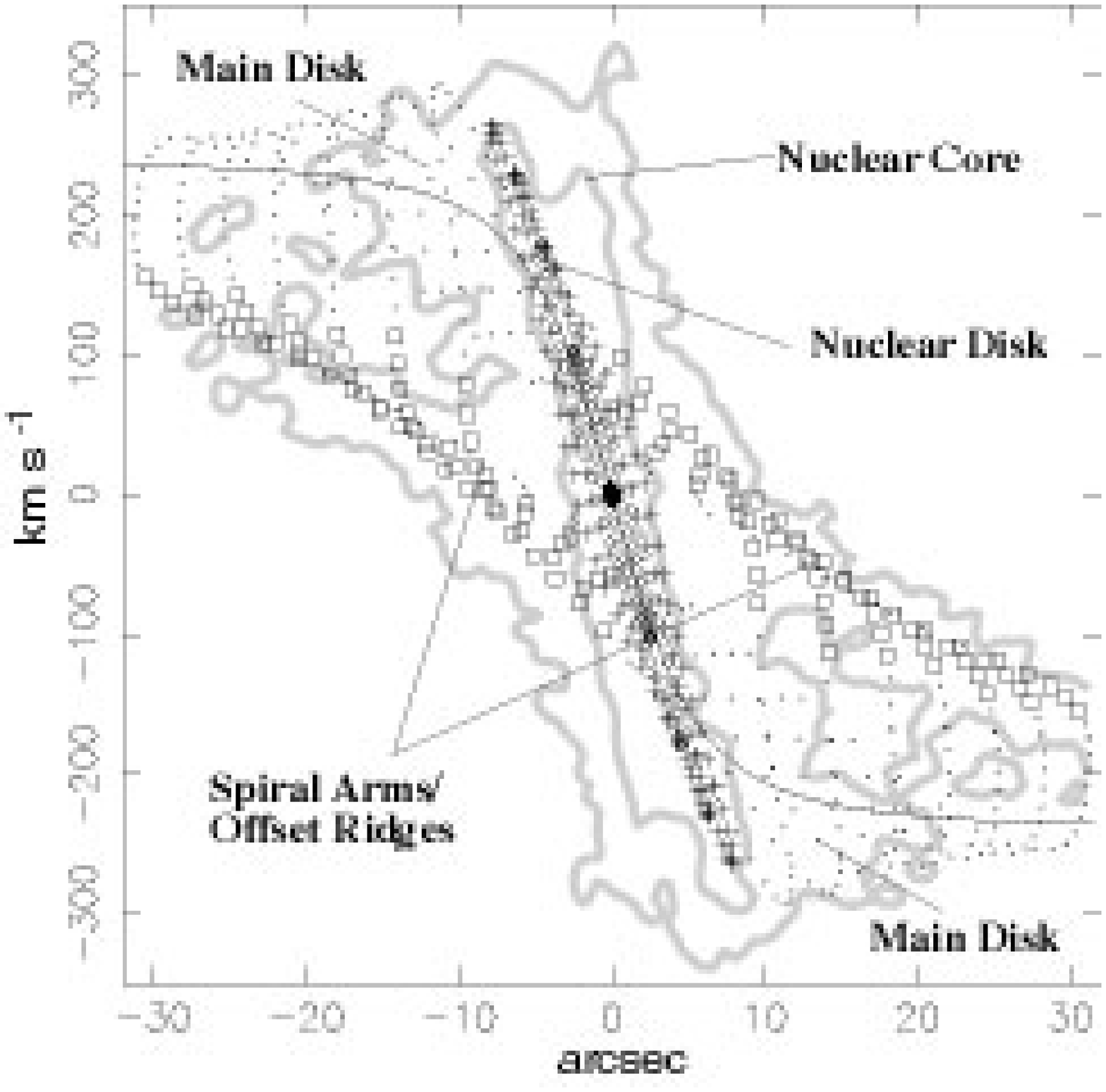}
\caption{
Comparison of the model and observed PV diagrams.
Contours trace the observed PV diagram (Figure \ref{fig:pvd}) at 8 and
40\% of the peak intensity. Symbols plotted are the same as those in
the bottom panel of Figure \ref{fig:orbits}. Our model reproduces
the main features of the molecular disk of NGC 3079
except the nuclear core. This indicates the presence of a massive
core.\label{fig:pvcomp}}
\end{figure}

\begin{figure}
\epsscale{1.0}
\plotone{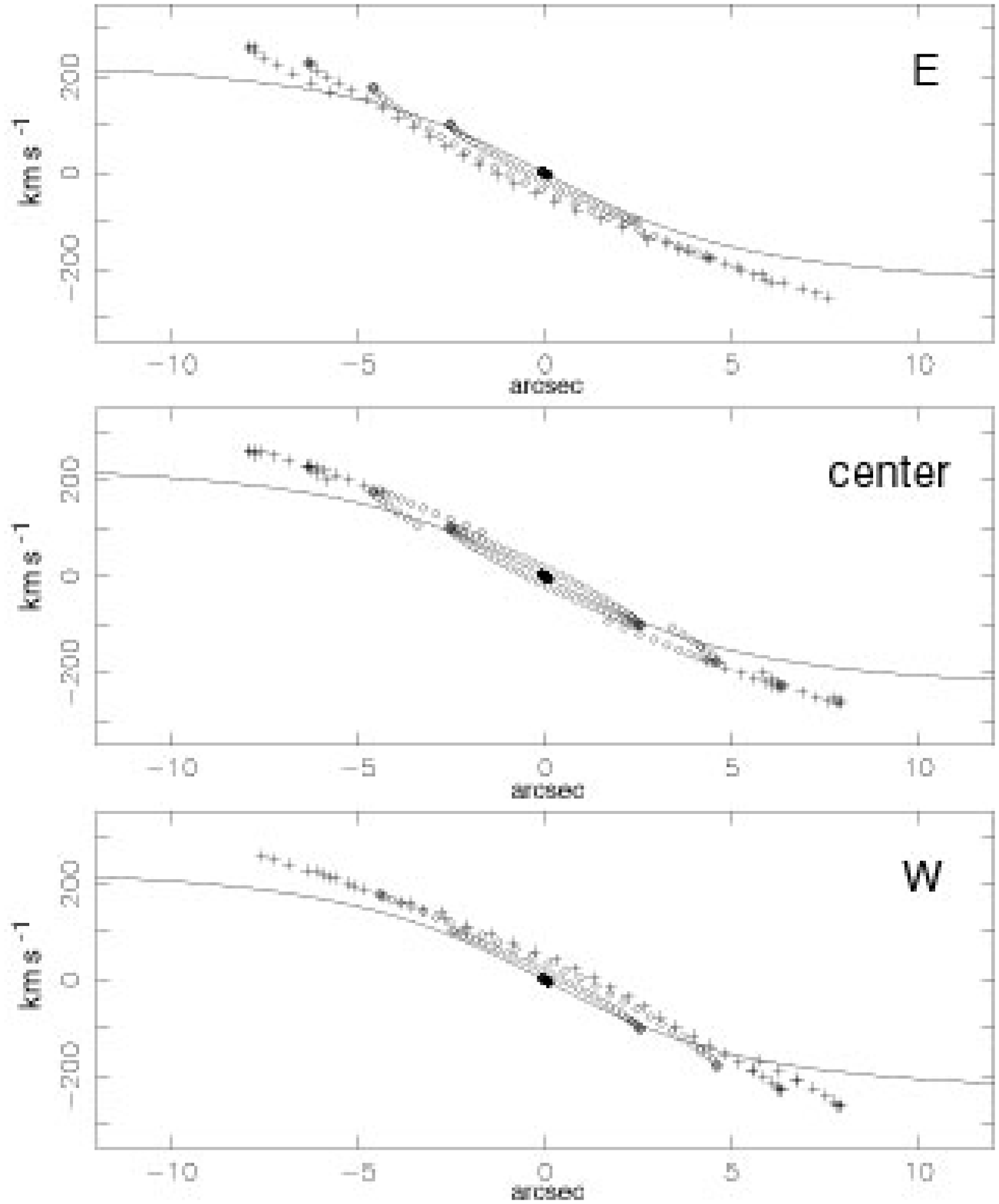}
\caption{
Model PV diagrams for the orbits within the outer ILR.
The diagrams are cut along the major axis for east ($y>0\arcsec$),
center and west ($y<0\arcsec$) sides, where $y$ is the vertical
axis on Figure \ref{fig:orbits} ({\it top}). The cut has a $10\arcsec$
width in the galaxy's face-on plane. The symbols are the same as
those in Figure \ref{fig:orbits} ({\it bottom}):
crosses are $x_2$-orbits, and circles are $x_1$-orbits inside the
inner ILR. The model well represents the bends (east and west) and
twist (center) discussed in \S \ref{sec:nstrm} (see Figure
\ref{fig:ewpvd}).\label{fig:ewmdlpvd}}
\end{figure}

\begin{figure}
\epsscale{1.0}
\plotone{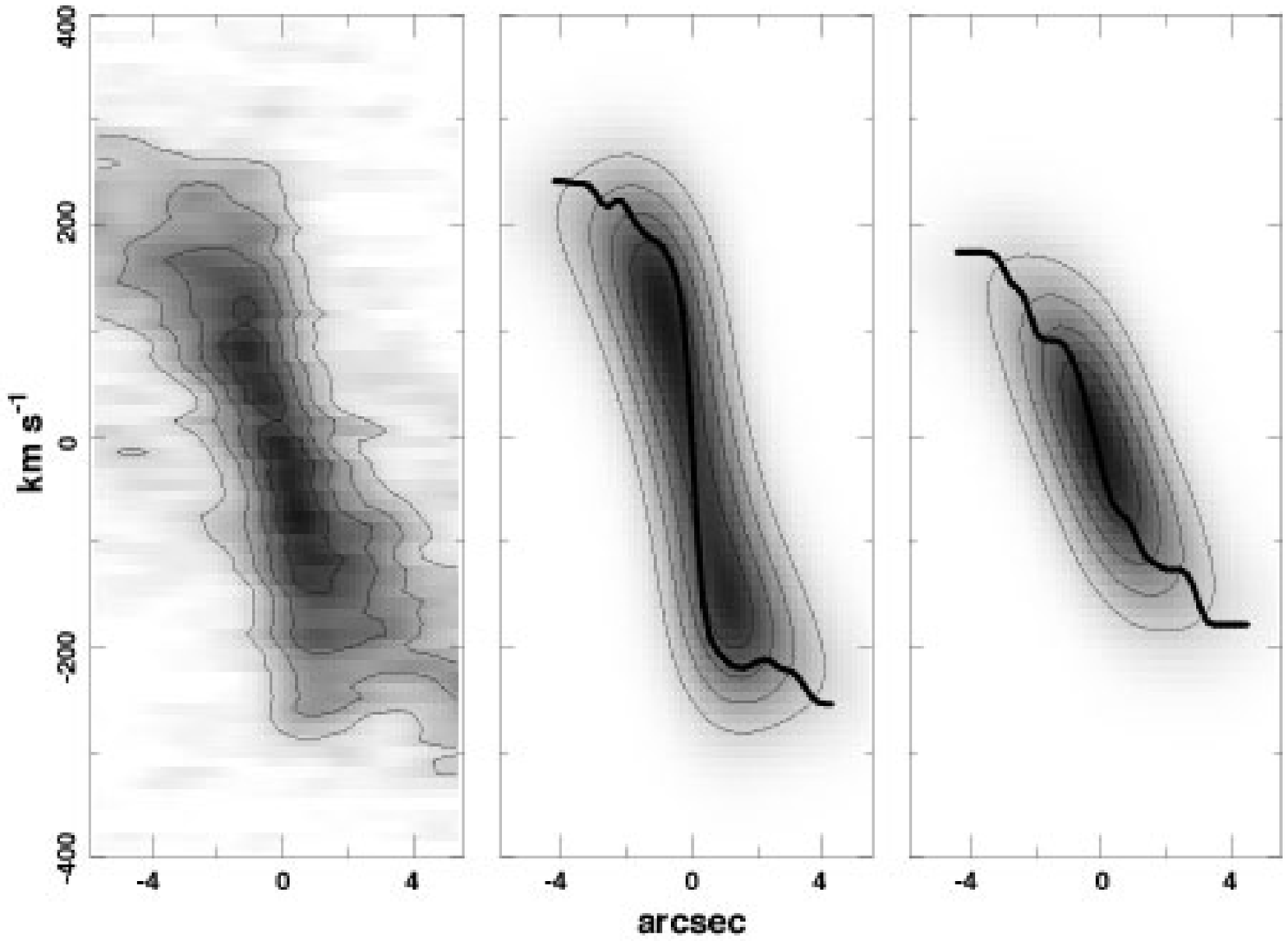}
\caption{PV diagrams and rotation curves. The slit is $1\arcsec$ wide.
Contours are 20, 40, 60, 80\% of the peak intensity in each panel.
{\it Left:} Observed PV diagram in the central region of NGC 3079.
{\it Middle:} PV diagram and rotation curve derived in the Takamiya
\& Sofue method.
{\it Right:} PV diagram and rotation curve derived in the peak-tracing
method.
\label{fig:tsrc}}
\end{figure}


\clearpage

\begin{thebibliography}{}

\bibitem[Achtermann \& Lacy(1995)]{al95} Achtermann, J. M. and Lacy, J. H.
   1995, \apj, 439, 163

\bibitem[Arimoto, Sofue and Tsujimoto(1996)]{ar96} Arimoto, N., Sofue, Y., and
   T. Tsujimoto 1996, \pasj, 48, 275

\bibitem[Athanassoula(1992)]{at92} Athanassoula, E. 1992, \mnras, 259, 345

\bibitem[Athanassoula \& Bureau(1999)]{ab99} Athanassoula, E. and Bureau, M.
   1999, \apj, 522, 699

\bibitem[Bally, Stark \& Wilson(1988)]{ba88} Bally, J., Stark, A. A., and
   Wilson, R. W. 1988, \apj, 324, 223

\bibitem[Binney et al.(1991)]{bn91} Binney, J. Gerhard, O. E., Stark, A. A.,
   Bally, J., and Uchida, K. I., 1991, \mnras, 252, 210

\bibitem[Braine et al.(1993)]{br93} Braine, J. et al. 1993, \aaps, 97, 887

\bibitem[Brandt(1965)]{br65} Brandt, J. C. 1965, \mnras, 129, 309

\bibitem[Bureau \& Athanassoula(1999)]{ba99} Bureau, M. and Athanassoula, E.
   1999, \apj, 522, 686

\bibitem[Cecil et al.(2001)]{ce01} Cecil, G., Bland-Hawthorn, J., Veilleux, S. and Filippeko, A. V.
   2001, \apj, 555, 338

\bibitem[Clemens, Sanders \& Scoville(1988)]{css88}
  Clemens, D. P., Sanders, D. B., and Scoville, N. Z. 1988, \apj, 327, 139

\bibitem[Combes et al.(1990)]{cm90} Combes, F., Debbasch, F., Friedli, D.,
   and Pfenniger, D. 1990, \aap, 233, 82

\bibitem[Condon(1997)]{co97} Condon, J. J. 1997, \pasp, 109, 166 

\bibitem[Cox et al.(2000)]{al00}
   Cox A. N. 2000, Allen's astrophysical quantities (4th ed), (New York: Springer-Verlag)

\bibitem[Dame, Hartmann and Thaddeus(2001)]{dm01} Dame, T. M., Hartmann, D.,
   and Thaddeus, P. 2001, \apj, 547, 792

\bibitem[de Vaucouleurs et al.(1991)]{rc3} de Vaucouleurs, G., de Vaucouleurs, A., Corwin, H. G.,
   Buta, R. J., Paturel, G., and Fouque, P. 1991, Third Reference Catalog of Bright Galaxies
   (New York: Springer-Verlag)

\bibitem[Duric \& Seaquist(1988)]{ds88} Duric, N. and Seaquist, E. R. 1988,
   \apj, 326, 574

\bibitem[Erwin \& Sparke(1999)]{es99} Erwin, P. and Sparke, L. S. 1999,
   \apjl, 521, 37
   
\bibitem[Fabbiano, Kim \& Trinchieri(1992)]{fb92} 
   Fabbiano, G., Kim, D. -W., and Trinchieri, G. 1992, \apjs, 80, 531

\bibitem[Filippenko \& Sargent(1992)]{fs92} Filippenko, A. V. and
   Sargent, W. L. W. 1992, \aj, 103, 28

\bibitem[Ford et al.(1986)]{fd86} Ford, H. C., Dahari, O., Jacoby, G. H.,
   Crane, P. C., and Ciardullo, R. 1986, \apjl, 311, 7

\bibitem[Friedli et al.(1996)]{fr96} Friedli, D., Wozniak, H., Rieke, M.,
   Martinet, L., and Bratschi, P. 1996, \aaps, 118, 461

\bibitem[Garc\'ia-Burillo \& Gu\'elin(1995)]{gb95} Garc\'ia-Burillo, S. and Gu\'elin, M.
   1995, \aap, 299, 657

\bibitem[Handa et al.(1990)]{hn90} Handa, T., Nakai, N, Sofue, Y.,
   Hayashi, M. and Fujimoto, M. 1990, \pasj, 42, 1

\bibitem[Handa et al.(1992)]{hn92} Handa, T., Sofue, Y., Ikeuchi, S.,
   Kawabe, R., and Ishizuki, S. 1992, \pasj, 44, L227

\bibitem[Hawarden et al.(1995)]{hw95} Hawarden, T. G., Israel, F. P.,
  Geballe, T. R., and Wade, R. 1995, \mnras, 276, 1197

\bibitem[Heckman(1980)]{hk80} Heckman, T. M. 1980, \aap, 87, 152

\bibitem[Heckman, Armus \& Miley(1990)]{hk90} Heckman, T. M., Armus, L.,
  and Miley, G. K. 1990, \apjs, 74, 833

\bibitem[Heller, Shlosman \& Englmaier(2001)]{hs01} Heller, C.,
  Shlosman, I., and Englmaier, P. 2001, \apj, 553, 661

\bibitem[Henkel et al.(1984)]{hk84} Henkel, C, G\"usten, R., Downes, D.,
  Thum, C., Wilson, T. L., and Biermann 1984, \aap, 141, L1

\bibitem[Ho, Fillipenko \& Sargent(1997)]{ho97} Ho, L. C., Fillipenko, A. V.,
  and Sargent, W. L. W. 1997, \apjs, 112, 315

\bibitem[Irwin \& Seaquist(1988)]{is88}
  Irwin, J. A. and Seaquist, E. R. 1988, \apj, 335, 658

\bibitem[Irwin \& Seaquist(1991)]{is91} Irwin, J. A. and Seaquist, E. R.
   1991, \apj, 371, 111

\bibitem[Irwin \& Sofue(1992)]{is92} Irwin, J. A., and Sofue, Y. 1992,
  \apjl, 396, 75

\bibitem[Ishihara et al.(2001)]{is01} Ishihara, Y., Nakai, N., Iyamoto, N.,
   Makishima, K., Diamond, P., and Peter, H. 2001, \pasj, 53, 215

\bibitem[Koda \& Wada(2002)]{kw02} Koda, J. and Wada, K. 2002, submitted to ApJ

\bibitem[Kohno et al.(2001)]{kn01} Kohno, K. et al.
   2001, in ASP Conf. Proc. 249, The Central kpc of Starbursts and AGN,
   ed. Knapen, J. H., Beckman, J. E., Shlosman, I., and Mahoney, T. J.
   (San Francisco: ASP), 672

\bibitem[Kuijken \& Merrifield(1995)]{km95} Kuijken, K. and Merrifield, M. R.
   1995, \apjl, 443, 13

\bibitem[Laine et al.(1990)]{ln90} Laine, S., Kenney, J. D. P., Yun, M. S.,
   and Gottesman, S. T. 1999, \apj, 511, 709

\bibitem[Lindblad \& Lindblad(1994)]{ll94}
  Lindblad, P. O. and Lindblad, P. A. B. 1994, in ASP Conf. Proc. 66,
  Physics of the Gaseous and Stellar Disks of the Galaxy,
  ed. King. I. R. (San Francisco: ASP), 29

\bibitem[Merrifield \& Kuijken(1999)]{mk99} Merrifield, M. R. and Kuijken, K.
   1999, \aap, 345, L47

\bibitem[Miyoshi et al.(1995)]{mi95} Miyoshi, M., Moran, J., Henstein, J.,
  Greenhill, L., Nakai, N., Diamond, P., and Inoue, M. 1995, \nat, 373, 127

\bibitem[Nakai et al.(1995)]{na95} Nakai, N., Inoue, M., Miyazawa, K.,
  Miyoshi, M., and Hall, P. 1995, \pasj, 47, 771

\bibitem[Okumura et al.(2000)]{ok00} Okumura, S. K., et al. 2000, \pasj, 52, 393

\bibitem[P\'erez Garc\'ia, Rodr\'iguez Espinosa \& Fuensalida(2000)]{pg00}
  P\'erez Garc\'ia, A. M., Rodr\'iguez Espinosa, J. M., and Fuensalida, J. J.
  2000, \apj, 529, 875

\bibitem[Pietcsh, Trinchieri \& Vogler(1998)]{pt98}
  Pietsch, W. Trinchieri, G., and Vogler, A. 1998, \aap, 340, 351

\bibitem[Piner, Stone \& Teuben(1995)]{pst95} Piner, B. G., Stone, J. M., and
   Teuben, P. J. 1995, \apj, 449, 508

\bibitem[Regan et al.(2001)]{re01} Regan, M. W., Thornley, M. D., Helfer, T. T.,
   Sheth, K., Wong, T., Vogel, S. N., and Bock, D. C. -J. 2001, \apj, 561, 218

\bibitem[Sakamoto et al.(1999a)]{sk99} Sakamoto, K., Okumura, S. K., Ishizuki, S.,
   and Scoville, N. Z. 1999a, \apjs, 124, 403

\bibitem[Sakamoto et al.(1999b)]{sk99b} Sakamoto, K., Okumura, S. K., Ishizuki, S.,
   and Scoville, N. Z. 1999b, \apj, 525, 691

\bibitem[Sakamoto, Barker \& Scoville(2000)]{sk00} Sakamoto, K., Baker, A. J., and Scoville, N. Z. 2000, \apj, 533, 149

\bibitem[Sandage \& Tamman(1981)]{st81} Sandage, A., and Tamman, G. 1981,
   A Revised-Sharpley-Ames Catalog of Bright Galaxies
   (Washington, DC: Carnegie Institution of Washington)

\bibitem[Sawada-Satoh et al.(2000)]{ss00} Sawada-Satoh, S., Inoue, M,
  Shibata, K. M., Kameno, S., and Migenes, V. 2000, \pasj, 52, 428

\bibitem[Shaw, Wilkinson \& Carter(1993)]{sw93} Shaw, M., Wilkinson, A.,
   and Carter, D. 1993, \aap, 268, 511

\bibitem[Shaw et al.(1995)]{sw95} Shaw, M. A., Axon, D. J., Probst, R.,
   and Gatley, I. 1995, \mnras, 274, 369

\bibitem[Sofue \& Irwin(1992)]{si92} Sofue, Y., and Irwin, J. A. 1992, \pasj, 44, 353

\bibitem[Sofue \& Rubin(2001)]{sr01} Sofue, Y. and Rubin, V. 2001, \araa, 39, 137

\bibitem[Sofue et al.(1999)]{sf99} Sofue, Y., Tutui, Y., Honma, M.,
  Tomita, A., Takamiya, T., Koda, J., and Takeda, Y. 1999, \apj, 523, 136

\bibitem[Sofue et al.(2001)]{sf01} Sofue, Y., Koda, J., Kohno, K.,
  Okumura, S. K., Honma, M., Kawamura, A., and Irwin, J. A.
  2001, \apjl, 547, 115

\bibitem[Soifer et al.(1989)]{so89} Soifer, B. T., Boehmer, L., Neugebauer, G., and Sanders, D. B. 1989
   \aj, 98, 766

\bibitem[Sosa-Brito, Tacconi-Garman \& Lehnert(2001)]{sb01}
  Sosa-Brito, R. M., Tacconi-Garman, L. E., and Lehnert, M. D.
  2001, \apjs, 136, 61

\bibitem[Takamiya \& Sofue(2000)]{ts00} Takamiya, T. and Sofue, Y. 2000,
  \apj, 534, 670

\bibitem[Trotter et al.(1998)]{tr98} Trotter, A. S., Greenhill, L. J.,
   Moran, J. M., Reid, M. J., Irwin, J. A., and Lo, K. Y. 1998,
   \apj, 495, 740

\bibitem[Veilleux et al.(1994)]{vl94} Veilleux, S., Cecil, G.,
  Bland-Hawthorn, J., Tully, R. B. 1994, \apj, 433, 48

\bibitem[Veilleux, Bland-Hawthorn \& Cecil(1999)]{vl99} Veilleux, S.,
   Bland-Hawthorn, J., and Cecil, G. 1999, \apj, 118, 2108

\bibitem[Wada(1994)]{wa94} Wada, K. 1994, \pasj, 46, 165

\bibitem[Wada \& Koda(2001)]{wk01} Wada, K. and Koda, J. 2001, \pasj, 53, 1163

\bibitem[Wandel, Peterson \& Malkan(1999)]{wa99}
  Wandel, A., Peterson, B. M., and Malkan, M. A. 1999, \apj, 526, 579

\bibitem[Young, Claussen \& Scoville(1988)]{ycs88} Young, J. S., Claussen, M. J.,
  and Scoville, N. Z. 1988, \apj, 324, 115

\bibitem[Young et al.(1995)]{yg95} Young, J. S., et al. 1995, \apjs, 98, 219

\bibitem[Zacharias et al.(2000)]{usno2} Zacharias, N., et al. 2000, \aj, 120, 2131

\end{thebibliography}
\end{document}